\documentclass[a4paper,12pt,oneside]{article}
\usepackage[latin1]{inputenc}

\usepackage{amsmath,amsfonts,amstext,amscd,bezier,amssymb}

\usepackage[top=2.0cm,bottom=2cm,left=1.5cm,right=2cm]{geometry}

\usepackage{cancel} 
\usepackage{graphicx}
\usepackage{color}
\usepackage{tabularx}
\usepackage{xifthen}
\usepackage{lastpage}
\usepackage{psfrag}
\usepackage{float}
\usepackage{subfigure}
\usepackage{float}
\usepackage{fancyhdr}
\usepackage{upgreek}
\usepackage{pifont} 
\usepackage{eso-pic} 
\usepackage{textcomp}
\usepackage{algorithm}
\usepackage{algpseudocode}
\usepackage[colorlinks=false,pagebackref=true]{hyperref}

\usepackage{mathrsfs} 

\usepackage{tikz}
\usetikzlibrary{shapes,arrows}

\usepackage{listings}

\newcommand{\R}{\text{$\mathbb R$}}

\newcommand{\dparcial}[2]{\dfrac{\partial #1}{\partial #2}}

\newcommand{\vb}{\mathbf{v}}

\newcommand{\qb}{\mathbf{q}}

\newcommand{\ds}{\displaystyle}

\usepackage{authblk}

{\hfill$\rule{2mm}{2mm}$\medskip}

\title{\textsc{Variational Time Integration Approach for Smoothed Particle Hydrodynamics
Simulation of Fluids}}

\author[1,2]{Leandro Tavares da Silva\thanks{leandrots@gmail.com}}
\author[1]{Gilson Antonio Giraldi\thanks{gilson@lncc.br}}
\affil[1]{National Laboratory for Scientific Computing,  Petr\'{o}polis 25651-075, Brazil.}
\affil[1,2]{Federal Center of Technology Education Celso Suckow da Fonseca, Petr\'{o}polis 25620-003, Brazil.}

\begin{document}
\maketitle
\begin{abstract}
Variational time integrators are derived in the context
of discrete mechanical systems. In this area, the governing equations
for the motion of the mechanical system are built following two steps:
(a) Postulating a discrete action; (b) Computing the stationary point
for the discrete action. The former is formulated by considering Lagrangian
(or Hamiltonian) systems with the discrete action being constructed
through numerical approximations of the action integral. The latter
derives the discrete Euler-Lagrange equations whose solutions give
the variational time integrator. In this paper, we build variational
time integrators in the context of smoothed particle hydrodynamics
(SPH). So, we start with a variational formulation of SPH for fluids.
Then, we apply the generalized midpoint rule, which depends on a parameter
$\alpha$, in order to generate the discrete action. Then, the step
(b) yields a variational time integration scheme that reduces to a
known explicit one if $\alpha\in\{0,1\}$ but it is implicit otherwise.
Hence, we design a fixed point iterative method to approximate the
solution and prove its convergence condition. Besides, we show that
the obtained discrete Euler-Lagrange equations preserve linear momentum.
In the experimental results, we consider artificial viscous as well as boundary
interaction effects and simulate a dam breaking set up. We compare
the explicit and implicit SPH solutions and analyze momentum conservation
of the dam breaking simulations.
\end{abstract}



\section{Introduction}

Fluid simulation involves numerous works that can be coarsely classified
in partial differential equations (PDEs) and lattice based techniques. PDEs methods derive computational
models based on continuous fluid equation, like the Navier-Stokes
ones, and numerical techniques formulated through discretization approaches
that can be Lagrangian (Smoothed Particle Hydrodynamics (SPH) \cite{Liu2003b}, Moving Particle \cite{Kim-Park2014},
Moving-Particle Semi-Implicit \cite{Ataie-Ashtiani2006}) or Eulerian (Finite Element, Finite Difference and Finite Volume) \cite{John-Wendt2009}. Lattice
based approaches are built using cellular automata and lattice Boltzmann methods \cite{Chopard1998b}.

In this paper we focus on the SPH technique, that was originally invented
to solve astrophysical problems in three dimensional open space \cite{Gingold77,Monaghan92}.
It is a meshfree, Lagrangian approach based on particle systems and
interpolation theory. Kinematic and dynamic variables, such as velocity,
density, deformation gradient and stresses are obtained from the fluid
flow at the particle positions using interpolation functions known
as kernels. Since its invention , SPH has been extensively studied
and extended to address scientific and engineering problems in material
science, free surface flows, explosion phenomena, heat transfer, mass
flow, among many other applications (see \cite{Monaghan2012Smoothed} and references
therein).

The dynamic model behind SPH is based on classical mechanics which
is concerned with physical laws to describe the behavior of a macroscopic
system under the action of forces \cite{goldstein_classical_1981}. For
instance, when considering a particle system in the $3D$ space under
the action of gravity, we can take the position vector of each particle
along the time $t$, which in cartesian coordinates is given by $\left(x^{i1},x^{i2},x^{i3}\right)\in\mathbb{R}^{3}$,
$i=1,2,\ldots,M$, and use the Newton's laws to get the governing
equations written in terms of the cartesian coordinates and the time
$t$. In a more general situation, the instantaneous configuration
of each particle may be described by the values of $n$ \emph{generalized}
coordinates $(q^{i1},q^{i2},\cdot\cdot\cdot,q^{in})$. So, we need a
methodology to write the evolution equations of the system in terms
of coordinates other than the cartesian ones.

The Lagrangian formulation of mechanics is a framework to address
this issue. It is a variational approach
based on the integral Hamilton's principle which states that the correct
path of the motion of a system is a stationary point for the action
integral \cite{goldstein_classical_1981}. The corresponding Lagrange's
equations allow to write the evolution of the system, the SPH fluid
particles in our case, in term of the generalized coordinates. Then,
we update the velocities and positions of the particles
by using a suitable time integrator. This methodology is
followed by the variational approaches for SPH simulation of fluids
found in the literature \cite{Bonet1999,Bonet2004}.

In this paper we follow a different variational formulation, based
on discrete mechanics concepts. The fundamental point of the theory
of discrete mechanics consists of discretizing Hamilton's principle
of Lagrangian mechanics \cite{Hairer2002}. Consequently, in discrete mechanics,
the time evolution of the mechanical system is obtained following
two steps: (a) Computing a discrete action; (b) Postulating that the
corresponding path is a stationary point for the discrete action.
The former is implemented by considering Lagrangian (or Hamiltonian)
systems with the discrete action being constructed through numerical
approximations of the action integral. The latter derives the discrete
Euler-Lagrange equations whose solutions give the variational time
integration technique \cite{Marsden-West2001}.

In this paper, we derive variational time integrators in the context
of SPH. Up to the best of our knowledge, such approach has not been
used by the SPH community before. So, we start with the continuum variational
formulation of SPH for fluids described in \cite{Bonet2004}. Then, we apply
the generalized midpoint rule in order to build the discrete action \cite{Lew-variational2003}.
This numerical integration rule depends on a parameter $\alpha$ which
is further explored in the text. So, we show that the discrete Euler-Lagrange
equations reduces to the known Verlet technique if $\alpha\in\{0,1\}$.
For $0<\alpha<1$ we obtain an implicit integration scheme. We demonstrate
the sufficient condition to apply the contraction mapping principle
and, consequently, to cast the implicit scheme as a method of successive
approximations to get the solution. Momentum conservation is also
demonstrated for the discrete Euler-Lagrange equations. In the implementation
details, numerical aspects and boundary interaction effects are added
to the discrete Euler-Lagrange equations. In the experimental results,
we simulate a $2D$ dam breaking set up. We compare the explicit ($\alpha\in\{0,1\}$)
and the implicit SPH solution obtained by $\alpha=0.5$, and analyze
momentum conservation of the time integrators.

The remainder of this paper is organized as follows. Section \ref{sec:related}
describes related works. The discrete Lagrangian mechanics is presented
on section \ref{sec:Lagragian}. Next, in section \ref{sec:Lagrange-SPH},
we describe the Lagrangian formulation of SPH. The derivation of the
variational time integrator for SPH is presented on section \ref{sec:SPH-MidPointRule}.
This section also demonstrates the momentum conservation of the obtained
discrete Euler-Lagrange equations (section \ref{subsec:Momentum-Conservation})
as well as the application of contraction mapping principle and implementation details (sections
\ref{subsec:Fixed-Point-SPH} and \ref{subsec:Implementation}). The computational results and conclusions/future
works are presented on sections \ref{sec:Computational-Experiments}
and \ref{sec:Conclusions-and-Future}, respectively.

\section{Related Works \label{sec:related}}

Variational integrators in mechanical systems start by considering
mechanics from a variational point of view, following remarkable works
of Lagrange and Hamilton \cite{goldstein_classical_1981}. The Hamilton's principle or the
least action principle allows to cast the Newton's framework into
a geometric viewpoint in which the path followed by the physical system
in the configuration space has optimal geometric properties analogously
to the notion of geodesics on curved surfaces \cite{Gaw2011}. Therefore, we can design
numerical integrators that exploit the geometric structure behind
mechanical systems, which are named geometric integrators \cite{Hairer2002,Gaw2011}.
A special class of geometric integrators, called variational integrators,
discretized the variational formulation of mechanics generating iterative
schemes to compute an approximation for the path of the physical system
with any order of accuracy. Besides, this discrete geometric
framework can handle constraints, external and dissipative forces
making variational integrators both versatile and powerful \cite{Kane-Marsden-Ortiz1999,Lew-variational2003,Lew-Marsden2004}.

In the Lagrangian point of view, given a mechanical system with configuration
space (manifold) $Q$, the Lagrangian itself is a real map defined
in the velocity phase space. The first step to represent the system
in discrete variational mechanics frameworks is to replace the velocity
phase space by $Q\times Q$ through some integration rule in order
to convert the continuous action in a discrete one. However, the Noether's
theorem allows to characterize the essence of a mechanical system
through its symmetries and invariants. Thus preserving these symmetries
and invariants into the discrete computational approaches is fundamental
to properly capture the correct continuous motion. In fact, it can
be shown that invariants can be preserved by variational time integrators
due to the fact that they respect variational nature of dynamics \cite{Lew-Marsden2004}.
This property, together with the fact that variational approaches
gives an unified view on both discrete mechanics and integration methods
for mechanical systems motivate the application of these frameworks
for computational models in solids \cite{Lew-variational2003, Mata-Lew2014}, optimal control \cite{Colombo-Zuccalli2013}, n-body problems
\cite{Lee-Leok-McClamroch2009}, computer animation \cite{Kharevych-Marsden-Desbrun2006,Mullen-Keenan-Yiying2009} and celestial mechanics \cite{Lee-Leok-McClamroch2007b}.

The key elements in variational time integrators are the discrete
action sum, the discrete Euler-Lagrange equations and the discrete
Noether's theorem that were clearly understood due to early works (see
\cite{Marsden-West2001} and references therein). Numerical aspects and convergence properties were specifically
considered in \cite{Kane-Marsden-Ortiz1999, Muller-Ortiz2004}.

On the other hand, traditional SPH formulations for fluids rely on
standard conservation equations and a particle framework to discretize
the corresponding Navier-Stokes equations, generating models that
treat the continuum fluid as a system of particles and recover continuous fields by using interpolation kernels \cite{Violeau2012Fluid}. Variational formulations
of SPH for fluid applications have been proposed, where the constitutive
equation of the fluid is given by an internal energy term which is
a function of the density \cite{Bonet1999,Bonet2004}. These formulations provide a basis
to discuss momentum preserving properties of SPH approaches. In this paper, they are used to derive variational time integrators
for SPH by computing the discrete action and its stationary point, as we shall see in the next sections.

\section{Discrete Lagrangian Mechanics \label{sec:Lagragian}}

Let us consider a physical system whose instantaneous configuration
may be described by the values of $n$ generalized coordinates $\mathbf{q}=(q^{1},q^{2},\cdot\cdot\cdot,q^{n})$
which is a point in a $n-dimensional$ Cartesian hyperspace known
as \emph{configuration space}. As time goes on from a time $t_{1}$
to a time $t_{2}$, the system changes its configuration due to internal
and external forces. Therefore, the evolution of the system can be
seem as a continuous path $\mathbf{q}(t)$, in the configuration space,
parameterized through the time $t$.

The Hamilton's principle gives a methodology to write the evolution
equation of the system in terms of the generalized coordinates and
time $t$. So, given the kinetic energy $K=K\left({\mathbf{\dot{q}}}\right),$
where ${\mathbf{\dot{q}}}=d\mathbf{q/}dt,$ and a scalar potential
$P\left(\mathbf{q},{\mathbf{\dot{q}}},t\right)=U\left(\mathbf{q}\right)+V\left(\mathbf{q,}{\mathbf{\dot{q}},t}\right)$,
where $U$ and $V$ accounts for conservative and non-conservative
velocity-dependent forces, the Hamilton's principle states that the
motion of the system from time $t_{1}$ to time $t_{2}$ is such that
the line integral:

\begin{equation}
S\left(\mathbf{q}\right)=\int_{t_{1}}^{t_{2}}L\left(\mathbf{q,}{\mathbf{\dot{q}}},t\right)dt,\label{Lagran00}
\end{equation}
where $L\left(\mathbf{q},{\mathbf{\dot{q}},t}\right)=T\left({\mathbf{\dot{q}}}\right)-U\left(\mathbf{q}\right)-V\left(\mathbf{q},{\mathbf{\dot{q}},t}\right)$,
named the Lagrangian of the system, has a stationary point for the
correct path of the motion \cite{goldstein_classical_1981}.

In discrete mechanics, we derive the governing equations of a physical
system by firstly considering a time sequence $t_{0},t_{1},\ldots,t_{N}$
to write: $\mathbf{q}\left(t_{0}\right)\equiv\mathbf{q}_{0}$, $\mathbf{q}\left(t_{1}\right)\equiv\mathbf{q}_{1}$,$\cdot\cdot\cdot$, $\mathbf{q}\left(t_{N}\right)\equiv\mathbf{q}_{N}$.

In this way, the system evolution is represented by a discrete trajectory
$\left(\mathbf{q}_{k},t_{k}\right),$ $k=0,1,\ldots,N,$ and the action
in equation (\ref{Lagran00}) becomes the \emph{discrete action},
given by:

\begin{equation}
S_{d}\left(\mathbf{q}_{0},\mathbf{q}_{1},\ldots,\mathbf{q}_{N}\right)=\sum_{k=0}^{N-1}L_{d}\left(\mathbf{q}_{k},\mathbf{q}_{k+1}\right),\label{eq:discrete-action00}
\end{equation}
where:

\begin{equation}
L_{d}\left(\mathbf{q}_{k},\mathbf{q}_{k+1}\right)\approx\int_{t_{k}}^{t_{k+1}}L\left(\mathbf{q,\dot{q},}t\right)dt.\label{eq:definition-Ld-in-action}
\end{equation}
is called the \emph{discrete Lagrangian}.

So, we can get a discrete version of the Hamilton's principle by considering
a family $\mathbf{q}_{k}\left(\varepsilon\right),$ $k=0,1,2,\ldots N$
such that $\mathbf{q}_{0}\left(\varepsilon\right)=\mathbf{q}_{0}$
and $\mathbf{q}_{N}\left(\varepsilon\right)=\mathbf{q}_{N}$, for
all $\varepsilon$ (end points fixed). So, in this case:

\[
S_{d}\left(\mathbf{q}_{0}\left(\varepsilon\right),\mathbf{q}_{1}\left(\varepsilon\right),\ldots,\mathbf{q}_{N}\left(\varepsilon\right)\right)=S_{d}\left(\varepsilon\right),
\]
and:

\begin{equation}
\delta S_{d}\equiv\left(\frac{dS_{d}\left(\varepsilon\right)}{d\varepsilon}\right)_{\varepsilon=0}=\sum_{i=0}^{N}\frac{\partial S_{d}}{\partial\mathbf{q_{i}}}\delta\mathbf{q}_{i},\label{eq:discrete-variation-00}
\end{equation}
where:

\[
\delta\mathbf{q}_{i}=\left(\frac{d\mathbf{q_{i}}}{d\varepsilon}\right)_{\varepsilon=0},
\]

Finally, analogously to the continuous case, we postulate that the
desired (discrete) path must satisfies $\delta S_{d}\left(\mathbf{q}\right)=0$,
which renders:

\begin{equation}
\frac{\partial}{\partial\mathbf{q}_{k}}L_{d}\left(\mathbf{q}_{k-1},\mathbf{q}_{k}\right)+\frac{\partial}{\partial\mathbf{q}_{k}}L_{d}\left(\mathbf{q}_{k},\mathbf{q}_{k+1}\right)=0,\label{eq:Discrete-ELE}
\end{equation}
for $k=1,2,\ldots,N-1$, which are the discrete Euler-Lagrange
equations \cite{Marsden-West2001}.

\section{Lagrangian Formulation for SPH \label{sec:Lagrange-SPH}}

The two fundamental elements in the SPH method are the interpolation
kernel $W:\mathbb{R}^{3}\rightarrow\mathbb{R}^{+}$, which is a symmetric
function respect to the origin $\left(0,0,0\right)$, bounded, with
compact support, and a particle system $\mathbf{q}^{i}=\left(x^{i1},x^{i2},x^{i3}\right)\in\mathbb{R}^{3}$,
$i=1,2,\ldots,M$, that represents a discrete version (samples) of
the fluid. The kernel estimate of a scalar quantity $A$ and its gradient
in a point $\mathbf{q}^{i}\in\mathbb{R}^{3}$ are given by \cite{Liu2003b}:
\begin{equation}
\left\langle A(\mathbf{q}^{i})\right\rangle =\sum_{j=1}^{M}\frac{m_{j}}{\rho(\mathbf{q}^{j})}A(\mathbf{q}^{j})W(\mathbf{q}^{i}-\mathbf{q}^{j},h),\label{sph1}
\end{equation}

\begin{equation}
\left\langle \mathbf{\nabla}A(\mathbf{q}^{i})\right\rangle =\sum_{j=1}^{M}\frac{m_{j}}{\rho(\mathbf{q}^{j})}A(\mathbf{q}^{j})\nabla_{i}W(\mathbf{q}^{i}-\mathbf{q}^{j},h),\label{eq:gradient-sph}
\end{equation}
where $\nabla_{i}W(\mathbf{q}^{i}-\mathbf{q}^{j},h)$ means $\nabla_{\mathbf{r}}W(\mathbf{r}-\mathbf{q}^{j},h)$
evaluated at $\mathbf{r}=\mathbf{q}^{i}$, $h$ is the smoothing length
which determines the support of the kernel and $\rho\left(\mathbf{q}^{j}\right)$ is the density at the particle
position $\mathbf{q}^{j}$ \cite{Liu2003b}.
Therefore, the kernel estimate of the density at the position $\mathbf{q}^{i}$
is:
\begin{equation}
\left\langle \rho\left(\mathbf{q}^{i}\right)\right\rangle = \sum_{j=1}^{M}m_{j}W(\mathbf{q}^{i}-\mathbf{q}^{j},h). \label{eq:density-sph00}
\end{equation}

Besides, we can show that the divergence of a vector field $\mathbf{v}$ can be computed
as \cite{Liu2003b}:
\begin{equation}
\left\langle \mathbf{\nabla\cdot}\mathbf{v}(\mathbf{q}^{i})\right\rangle =\sum_{j=1}^{M}\frac{m_{j}}{\rho(\mathbf{q}^{j})}\left(\mathbf{v}\left(\mathbf{q}^{j}\right)- \mathbf{v}\left(\mathbf{q}^{i}\right)\right)\nabla_{i}W(\mathbf{q}^{i}-\mathbf{q}^{j},h). \label{eq:sph-divergent}
\end{equation}

For simplicity, in what follows, we take off the brackets in the left hand
side of expressions (\ref{sph1})-(\ref{eq:density-sph00}). In this
work, the kernel function adopted is the Gaussian one:

\begin{equation}
W\left(R\right)=\ell e^{-R^{2}},\label{eq:Gaussian-Kernel}
\end{equation}
where $\ell$ is a constant. Also, in the SPH framework it is usually postulated a state equation that correlates density
and pressure, which in this work is given by:

\begin{equation}
p(\mathbf{q}^{i})=B\left[\left(\dfrac{\rho\left(\mathbf{q}^{i}\right)}{\rho_{0}}\right)^{7}-1\right],\label{eq:state-equation-sph}
\end{equation}
where $\rho_{0}$ is the rest density, $B$ is a constant such that $B=c^2\rho_0/7$, and $c$ is the speed sound in fluid.

The SPH model for a fluid can be seen as a system composed
by particles $\mathbf{q}^{i}=\left(x^{i1},x^{i2},x^{i3}\right)$ subject
to forces derived from internal and external potentials that are functions of fluid fields like density $\rho$, pressure
$p$, and velocity $\mathbf{v}$.
Moreover, the configuration of the SPH system along the time is
described by a continuous path in the configuration space:

\begin{equation}
\mathbf{q}\left(t\right)=\left(\left\{ \mathbf{q}^{1}\left(t\right)\right\} ^{T},\left\{ \mathbf{q}^{2}\left(t\right)\right\} ^{T},\cdot\cdot\cdot,\left\{ \mathbf{q}^{M}\left(t\right)\right\} ^{T}\right)^{T}\in\mathbb{R}^{3M}.\label{eq:sph-fluid-configuration}
\end{equation}

Such viewpoint is behind the (continuum) variational formulation of
SPH presented in \cite{Bonet1999,Bonet2004}. The total kinetic energy of
the system can be simply computed as the sum of the kinetic energy
of the particles:
\begin{equation}
K(\mathbf{q})=\dfrac{1}{2}\sum_{i=1}^{M}m_{i}(\mathbf{\dot{q}}^{i}\cdot\mathbf{\dot{q}}^{i})\label{eq:kinetic-sph}
\end{equation}

The potential energy is the sum of the external and internal potential energies:
\begin{equation}
P(\mathbf{q})=\Pi_{\text{ext}}+\Pi_{\text{int}}. \label{eq:total-pontential-sph}
\end{equation}
Thus, for the case where the external forces result from a gravitational
field $\mathbf{g}$, the total external energy is:
\begin{equation}
\Pi_{\text{ext}}=-\sum_{i=1}^{M}m_{i}(\mathbf{q}^{i}\cdot\mathbf{g}). \label{eq:external-potential-sph}
\end{equation}
On the other hand, the internal energy will incorporate the constitutive
characteristics of the system. In general,
it is possible to express the total internal energy as the sum:

\begin{equation}
\Pi_{\text{int}}=\sum_{i=1}^{M} m_{i} \pi(\rho\left(\mathbf{q}^{i}\right),\cdots)\label{eq:internal-potential-sph}
\end{equation}
where $\pi$ will depend on the deformation, density or other constitutive
parameters. In \cite{Bonet2004} the constitutive equations for
a nearly incompressible flow without dissipative effects is:

\begin{equation}
\frac{d\pi}{d\rho}=\frac{p}{\rho^{2}}, \label{eq:constitutive-equations-sph}
\end{equation}
where $p$ is the fluid pressure. In \cite{Bonet1999,Bonet2004} expressions (\ref{eq:kinetic-sph})-(\ref{eq:constitutive-equations-sph}) are used to compute
the Lagrangian:

\begin{equation}
L=K-\Pi_{\text{int}}-\Pi_{\text{ext}},\label{eq:Continuous-Lagrangean-SPH}
\end{equation}
and the governing equations of the SPH system of particles can be
yielded through the (continuous) Euler-Lagrange equations. Instead, in this work we follow a discrete approach described next.

\section{SPH Variational Time Integrator\label{sec:SPH-MidPointRule}}

To derive the discrete variational formulation for SPH systems we
need to build a discrete Lagrangian through expression (\ref{eq:definition-Ld-in-action})
and then insert the result in the discrete Euler-Lagrange equations
(\ref{eq:Discrete-ELE}). Following section \ref{sec:Lagragian},
we consider a time sequence $t_{0},t_{1},\ldots,t_{N}$ and a corresponding
discrete path of the SPH system in the configuration space, given
by:

\[
\mathbf{q}\left(t_{0}\right)=(\left\{ \mathbf{q}^{1}\left(t_{0}\right)\right\} ^{T},\left\{ \mathbf{q}^{2}\left(t_{0}\right)\right\} ^{T},\cdot\cdot\cdot,\left\{ \mathbf{q}^{M}\left(t_{0}\right)\right\} ^{T})^{T}\equiv\mathbf{q}_{0},
\]

\[
\mathbf{q}\left(t_{1}\right)=(\left\{ \mathbf{q}^{1}\left(t_{1}\right)\right\} ^{T},\left\{ \mathbf{q}^{2}\left(t_{1}\right)\right\} ^{T},\cdot\cdot\cdot,\left\{ \mathbf{q}^{M}\left(t_{1}\right)\right\} ^{T})^{T}\equiv\mathbf{q}_{1},
\]

\[
\ldots
\]

\[
\mathbf{q}\left(t_{N}\right)=(\left\{ \mathbf{q}^{1}\left(t_{N}\right)\right\} ^{T},\left\{ \mathbf{q}^{2}\left(t_{N}\right)\right\} ^{T},\cdot\cdot\cdot,\left\{ \mathbf{q}^{M}\left(t_{N}\right)\right\} ^{T})^{T}\equiv\mathbf{q}_{N}.
\]

Moreover, a numerical integration rule is necessary to approximate
the action in the interval $\left[t_{k},t_{k+1}\right]$. In this
work we choose the generalized midpoint rule which gives \cite{Marsden-West2001}:
\begin{eqnarray*}
L_{d}(\qb_{k},\qb_{k+1}) & = & (t_{k+1}-t_{k})L\left((1-\alpha)\qb_{k}+\alpha\qb_{k+1},\dfrac{\qb_{k+1}-\qb_{k}}{t_{k+1}-t_{k}}\right)\\
\end{eqnarray*}

\[
=(t_{k+1}-t_{k})\left[\frac{1}{2}\sum_{i}m_{i}\Biggl\|\dfrac{\qb_{k+1}^{i}-\qb_{k}^{i}}{t_{k+1}-t_{k}}\Biggl\|_{2}^{2}\right]
\]

\begin{equation}
-(t_{k+1}-t_{k})\left[\sum_{i}m_{i}\pi(\rho_{k,k+1}^{i})-\sum_{i}m_{i}((1-\alpha)\qb_{k}^{i}+\alpha\qb_{k+1}^{i})\cdot\mathbf{g}\right],\label{eq:Discrete-Lagrangian-1}
\end{equation}
where the Lagrangian $L$ is defined by expression (\ref{eq:Continuous-Lagrangean-SPH}),
the parameter $\alpha\in\left[0,1\right]$, and:

\[
\rho_{k,k+1}^{i}=\rho((1-\alpha)\qb_{k}^{i}+\alpha\qb_{k+1}^{i})
\]

\begin{equation}
=\sum_{j}m_{j}W\left(\beta_{k,k+1}^{i,j}\right)\label{eq:formula-gilson00},
\end{equation}
where:
\begin{equation}
\beta_{k,k+1}^{i,j}=(1-\alpha)\qb_{k}^{i}+\alpha\qb_{k+1}^{i}-\left((1-\alpha)\qb_{k}^{j}+\alpha\qb_{k+1}^{j}\right). \label{eq:Definition-of-point}
\end{equation}

So,
\[
\dparcial{L_{d}}{\qb_{k}^{i}}(\qb_{k},\qb_{k+1})
\]

\begin{equation}
=(t_{k+1}-t_{k})\left[m_{i}\left(\dfrac{\qb_{k+1}^{i}-\qb_{k}^{i}}{t_{k+1}-t_{k}}\right)\dfrac{-1}{t_{k+1}-t_{k}}-\dparcial{}{\qb_{k}^{i}}\sum_{\upsilon}m_{\upsilon}\pi(\rho_{k,k+1}^{\upsilon})+m_{i}(1-\alpha)\mathbf{g}.\right]\label{eq:derivative-Ld-resp-qk}
\end{equation}

However, through the Chain rule and the kernel $W$ definition in expression
(\ref{eq:Gaussian-Kernel}), we can show that:

\begin{equation}
\frac{\partial}{\partial\mathbf{q}_{k}^{i}}W\left(\beta_{k,k+1}^{i,j}\right)=-2\ell e^{-R^{2}}\left[(1-\alpha)\mathbf{q}_{k}^{i}+\alpha\mathbf{q}_{k+1}^{i}-\left((1-\alpha)\mathbf{q}_{k}^{j}+\alpha\mathbf{q}_{k+1}^{j}\right)\right](1-\alpha).\label{eq:newdw00}
\end{equation}

To simplify the equations in the remaining of this section we use
the notation:

\begin{equation}
\nabla_{i}W\left(\beta_{k,k+1}^{i,j}\right)\equiv-2\ell e^{-R^{2}}\left[(1-\alpha)\mathbf{q}_{k}^{i}+\alpha\mathbf{q}_{k+1}^{i}-\left((1-\alpha)\mathbf{q}_{k}^{j}+\alpha\mathbf{q}_{k+1}^{j}\right)\right]\label{eq:derivW00}
\end{equation}

Hence, by using the Chain rule, the constitutive equation (\ref{eq:constitutive-equations-sph})
involving pressure and density, and equation (\ref{eq:gradient-sph}),
we can prove that:

\[
\dparcial{}{\qb_{k}^{i}}\sum_{\upsilon}m_{\upsilon}\pi(\rho_{k,k+1}^{\upsilon})
\]

\begin{equation}
=\sum_{j}m_{i}m_{j}\left(\dfrac{p_{k,k+1}^{i}}{\left(\rho_{k,k+1}^{i}\right)^{2}}+\dfrac{p_{k,k+1}^{j}}{\left(\rho_{k,k+1}^{j}\right)^{2}}\right)\nabla_{i}W(\beta_{k,k+1}^{i,j})(1-\alpha),\label{eq:midpoint-pressure-force}
\end{equation}
where:

\begin{equation}
p_{k,k+1}^{i}=p((1-\alpha)\qb_{k}^{i}+\alpha\qb_{k+1}^{i}),\label{eq:formula-gilson-01}
\end{equation}

\begin{equation}
p_{k,k+1}^{j}=p\left((1-\alpha)\qb_{k}^{j}+\alpha\qb_{k+1}^{j}\right),\label{eq:formula-gilson02}
\end{equation}

Therefore, by inserting expression (\ref{eq:midpoint-pressure-force})
into equation (\ref{eq:derivative-Ld-resp-qk}) we obtain:

\[
\dparcial{L_{d}}{\qb_{k}^{i}}(\qb_{k},\qb_{k+1})
\]

\[
=-m_{i}\left(\dfrac{\qb_{k+1}^{i}-\qb_{k}^{i}}{\Delta t}\right)-\Delta t\sum_{j}m_{i}m_{j}\left(\dfrac{p_{k,k+1}^{i}}{\left(\rho_{k,k+1}^{i}\right)^{2}}+\dfrac{p_{k,k+1}^{j}}{\left(\rho_{k,k+1}^{j}\right)^{2}}\right)\nabla_{i}W(\beta_{k,k+1}^{i,j})(1-\alpha)
\]

\begin{equation}
+\Delta tm_{i}(1-\alpha)\mathbf{g},\label{eq:derivative-Ld-resp-qk-end}
\end{equation}
where $\Delta t=(t_{k+1}-t_{k})=constant$, $\beta_{k,k+1}^{i,j}$,
$p_{k,k+1}^{i}$ and $p_{k,k+1}^{j}$, are defined in expressions
(\ref{eq:Definition-of-point}), (\ref{eq:formula-gilson-01}) and
(\ref{eq:formula-gilson02}), respectively.\\

In the same way, we can calculate the action in the interval $\left[t_{k-1},t_{k}\right]$
to obtain:
\begin{eqnarray*}
L_{d}(\qb_{k-1},\qb_{k}) & = & \Delta tL\left((1-\alpha)\qb_{k-1}+\alpha\qb_{k},\dfrac{\qb_{k}-\qb_{k-1}}{\Delta t}\right)\\
\end{eqnarray*}

\begin{equation}
=\Delta t\left[\frac{1}{2}\sum_{i}m_{i}\left(\dfrac{\qb_{k}^{i}-\qb_{k-1}^{i}}{\Delta t}\right)^{2}-\sum_{i}m_{i}\pi(\rho_{k-1,k}^{i})+\sum_{i}m_{i}((1-\alpha)\qb_{k-1}^{i}+\alpha\qb_{k}^{i})\cdot\mathbf{g}\right],\label{eq:Action-in-the-tk-1-tk}
\end{equation}
where:

\[
\rho_{k-1,k}^{i}=\rho\left((1-\alpha)\qb_{k-1}^{i}+\alpha\qb_{k}^{i}\right)
\]
\begin{equation}
=\sum_{j}m_{j}W\left(\beta_{k-1,k}^{i,j}\right).\label{eq:density-for-k-k-1}
\end{equation}
where:
\begin{equation}
\beta_{k-1,k}^{i,j}=(1-\alpha)\qb_{k-1}^{i}+\alpha\qb_{k}^{i}-\left((1-\alpha)\qb_{k-1}^{j}+\alpha\qb_{k}^{j}\right),\label{eq:Definition-of-two-points}
\end{equation}

Then, analogously to expression (\ref{eq:newdw00}) we can demonstrate
that:

\begin{equation}
\frac{\partial}{\partial\mathbf{q}_{k}^{i}}W\left(\beta_{k-1,k}^{i,j}\right)=\nabla_{i}W\left(\beta_{k-1,k}^{i,j}\right)\alpha,\label{eq:newdw01}
\end{equation}
where $\nabla_{i}W\left(\beta_{k-1,k}^{i,j}\right)$ is computed by:

\begin{equation}
\nabla_{i}W\left(\beta_{k-1,k}^{i,j}\right) \equiv -2\ell e^{-R^{2}}\left[(1-\alpha)\qb_{k-1}^{i}+\alpha\qb_{k}^{i}-\left((1-\alpha)\qb_{k-1}^{j}+\alpha\qb_{k}^{j}\right)\right]\label{eq:derivW003}
\end{equation}

Then, using expressions (\ref{eq:Definition-of-two-points})-(\ref{eq:derivW003})
and following a development similar to the one performed to yield expression
(\ref{eq:derivative-Ld-resp-qk-end}) we can obtain:

\[
\dparcial{L_{d}}{\qb_{k}^{i}}(\qb_{k-1},\qb_{k})=-m_{i}\left(\dfrac{\qb_{k}^{i}-\qb_{k-1}^{i}}{\Delta t}\right)
\]

\begin{equation}
-\Delta t\sum_{j}m_{i}m_{j}\left(\dfrac{p^{i}_{k-1,k}}{\left(\rho^{i}_{k-1,k}\right)^{2}}+\dfrac{p^{j}_{k-1,k}}{\left(\rho^{j}_{k-1,k}\right)^{2}}\right)\nabla_{i}W(\beta_{k-1,k}^{i,j})\alpha+\Delta tm_{i}\alpha\mathbf{g},\label{eq:derivative-Ld-resp-qk-end-2}
\end{equation}
where:

\begin{equation}
p_{k-1,k}^{i}=p((1-\alpha)\qb_{k-1}^{i}+\alpha\qb_{k}^{i}),\label{eq:formula-gilson-01-1}
\end{equation}

\begin{equation}
p_{k-1,k}^{j}=p\left(\left((1-\alpha)\qb_{k-1}^{j}+\alpha\qb_{k}^{j}\right)\right).\label{eq:formula-gilson02-1}
\end{equation}

Now, if we insert expressions (\ref{eq:derivative-Ld-resp-qk-end})
and (\ref{eq:derivative-Ld-resp-qk-end-2}) into equation (\ref{eq:Discrete-ELE})
and re-arrange the terms we get:

\[
\dfrac{\qb_{k+1}^{i}-2\qb_{k}^{i}+\qb_{k-1}^{i}}{\Delta t}
\]

\[
=\Delta t(1-\alpha)\left[-\sum_{j}m_{j}\left(\dfrac{p_{k,k+1}^{i}}{\left(\rho_{k,k+1}^{i}\right)^{2}}+\dfrac{p_{k,k+1}^{j}}{\left(\rho_{k,k+1}^{j}\right)^{2}}\right)\nabla_{i}W(\beta_{k,k+1}^{i,j})+\mathbf{g}\right]
\]

\begin{equation}
+\Delta t\alpha\left[-\sum_{j}m_{j}\left(\dfrac{p_{k-1,k}^{i}}{\left(\rho_{k-1,k}^{i}\right)^{2}}+\dfrac{p_{k-1,k}^{j}}{\left(\rho_{k-1,k}^{j}\right)^{2}}\right)\nabla_{i}W(\beta_{k-1,k}^{i,j})+\mathbf{g}\right],\label{eq:ELdiscreto1}
\end{equation}
which defines the variational time integration scheme for SPH using
the generalized midpoint rule. This numerical scheme is an implicit
one, except for $\alpha\in\left\{ 0,1\right\} $, when it reduces
to the known Verlet technique.

To confirm this, let us set $\alpha=0$ in expression (\ref{eq:ELdiscreto1}).
From equation (\ref{eq:formula-gilson-01}) we shall notice that $p_{k,k+1}^{i}=p(\qb_{k}^{i})$,
if $\alpha=0$. Besides, if we set $\alpha=1$ in equation (\ref{eq:formula-gilson-01-1})
we get also $p_{k-1,k}^{i}=p(\qb_{k}^{i})$. Analogous results are
obtained for $\rho_{k,k+1}^{i}$ and $\rho_{k-1,k}^{i}$ in expressions
(\ref{eq:formula-gilson00}), (\ref{eq:density-for-k-k-1}), respectively.
As a consequence, we obtain the same explicit integration scheme for
both $\alpha=1$ and $\alpha=0$ in expression (\ref{eq:ELdiscreto1}),
given by:
\[
\ds\dfrac{\qb_{k+1}^{i}-2\qb_{k}^{i}+\qb_{k-1}^{i}}{\Delta t}
\]
\begin{equation}
=\Delta t\left[-\sum_{j}m_{j}\left(\dfrac{p(\qb_{k}^{i})}{\left(\rho(\qb_{k}^{i})\right)^{2}}+\dfrac{p(\qb_{k}^{j})}{\left(\rho(\qb_{k}^{j})\right)^{2}}\right)\nabla_{i}W(\qb_{k}^{i}-\qb_{k}^{j})+\mathbf{g}\right].\label{verlet}
\end{equation}

\subsection{Momentum Conservation\label{subsec:Momentum-Conservation}}

In the absence of external and dissipative forces the total linear momentum of a mechanical
system is preserved. We can use the framework of the discrete Noether's
Theorem to show that the integration scheme defined by equation (\ref{eq:ELdiscreto1})
meets this requirement \cite{Lew-variational2003}. On the other hand, we can follow
a more direct approach, and use expression (\ref{eq:ELdiscreto1})
to make explicit the relationship between the momentum variation of
a particle with mass $m_{i}$ and the internal forces:

\[
m_{i}\left(\dfrac{\qb_{k+1}^{i}-2\qb_{k}^{i}+\qb_{k-1}^{i}}{\left(\Delta t\right)^{2}}\right)=
\]

\begin{eqnarray}
 &  & (1-\alpha)\left[-\sum_{j}m_{i}m_{j}\left(\dfrac{p_{k,k+1}^{i}}{\left(\rho_{k,k+1}^{i}\right)^{2}}+\dfrac{p_{k,k+1}^{j}}{\left(\rho_{k,k+1}^{j}\right)^{2}}\right)\nabla_{i}W(\beta_{k,k+1}^{i,j})\right]\nonumber \\
 & + & \alpha\left[-\sum_{j}m_{i}m_{j}\left(\dfrac{p_{k-1,k}^{i}}{\left(\rho_{k-1,k}^{i}\right)^{2}}+\dfrac{p_{k-1,k}^{j}}{\left(\rho_{k-1,k}^{j}\right)^{2}}\right)\nabla_{i}W(\beta_{k-1,k}^{i,j}).\right],\label{eq:ELdiscreto1-1}
\end{eqnarray}
with $\beta_{k,k+1}^{i,j}$  and $\beta_{k-1,k}^{i,j}$ given by
equations (\ref{eq:Definition-of-point}) and (\ref{eq:Definition-of-two-points}),
respectively.

Due to the kernel properties \cite{Liu-Liu2003} we can show that:

\[
\nabla_{i}W(\beta_{k,k+1}^{i,j})=-\nabla_{j}W(\beta_{k,k+1}^{j,i}),\quad\nabla_{i}W(\beta_{k-1,k}^{i,j})=-\nabla_{j}W(\beta_{k-1,k}^{j,i}).
\]

By inserting these expressions in equation (\ref{eq:ELdiscreto1-1})
it is straightforward to show that:

\[
\sum_{i}m_{i}\left(\dfrac{\qb_{k+1}^{i}-2\qb_{k}^{i}+\qb_{k-1}^{i}}{\left(\Delta t\right)^{2}}\right)\equiv\sum_{i}m_{i}\mathbf{a}^{i}=\mathbf{0},
\]
which proves the preservation of linear momentum under the action
of internal forces.

In order to preserve the angular momentum we need more considerations.
Specifically, the discrete Noether's Theorem states that if the discrete
Lagrangian $L_{d}$ is invariant under the action of a transformation
group, then we have conservation of the associated momentum. In our
case, $L_{d}$ is given by expression (\ref{eq:Discrete-Lagrangian-1})
and we shall discard the external (gravitational) field for this analysis.
It is easy to show that the part of $L_{d}$ that accounts for the
kinetic energy is invariant under rotations, which is the transformation
group related to angular momentum. However, we need to apply specific
corrections in the traditional SPH kernels and/or gradient in order to fulfill this
invariance for the internal energy, as demonstrated in \cite{Bonet1999}. We
are not considering such corrections in this paper and, consequently,
we can not assure angular momentum conservation.

\subsection{Fixed Point Method \label{subsec:Fixed-Point-SPH}}

\begin{sloppypar} In this section we re-write equation (\ref{eq:ELdiscreto1})
as $\qb_{k+1}^{i}=\mathbf{F}^{i}\left(\qb_{k+1},\qb_{k},\qb_{k-1}\right)$
and we suppose that $\qb_{k}$, $\qb_{k-1}$ are known. Therefore,
we have $\mathbf{F}:\mathbb{R}^{3M}\rightarrow\mathbb{R}^{3M}$, where
$\mathbf{F}\left(\qb_{k+1},\qb_{k},\qb_{k-1}\right)=\left(\mathbf{F}^{1}\left(\qb_{k+1},\qb_{k},\qb_{k-1}\right),\mathbf{F}^{2}\left(\qb_{k+1},\qb_{k},\qb_{k-1}\right),\cdot\cdot\cdot,\mathbf{F}^{M}\left(\qb_{k+1},\qb_{k},\qb_{k-1}\right)\right)$
and we can seek for conditions for which $\mathbf{F}$ becomes a contraction
mapping respect to the unknown $\qb_{k+1}^{i}$. In this way, we can
find the solution of equation (\ref{eq:ELdiscreto1}) through a fixed
point iterative algorithm that is simpler to implement than the traditional
Newton's method \cite{Marsden-West2001}.\end{sloppypar}

Thus, from Equation (\ref{eq:ELdiscreto1}) we verified that:
\begin{equation}
\mathbf{F}^{i}\left(\qb_{k+1},\qb_{k},\qb_{k-1}\right)=2\qb_{k}^{i}-\qb_{k-1}^{i}+\left(\Delta t\right)^{2}(1-\alpha)\mathbf{H}_{g}^{i}\left(\qb_{k+1},\qb_{k}\right)+\left(\Delta t\right)^{2}\alpha\mathbf{H}_{g}^{i}\left(\qb_{k},\qb_{k-1}\right),\label{eq:derfinicao-F0}
\end{equation}
where:
\begin{eqnarray}
\mathbf{H}_{g}^{i}\left(\qb_{k+1},\qb_{k}\right) & = & \left[-\sum_{j}m_{j}\left(\dfrac{p_{k,k+1}^{i}}{\left(\rho_{k,k+1}^{i}\right)^{2}}+\dfrac{p_{k,k+1}^{j}}{\left(\rho_{k,k+1}^{j}\right)^{2}}\right)\nabla_{i}W(\beta_{k,k+1}^{i,j})+\mathbf{g}\right],\label{eq:derfinicao-F1}\\
\nonumber \\
\mathbf{H}_{g}^{i}\left(\qb_{k},\qb_{k-1}\right) & = & \left[-\sum_{j}m_{j}\left(\dfrac{p_{k-1,k}^{i}}{\left(\rho_{k-1,k}^{i}\right)^{2}}+\dfrac{p_{k-1,k}^{j}}{\left(\rho_{k-1,k}^{j}\right)^{2}}\right)\nabla_{i}W(\beta_{k-1,k}^{i,j})+\mathbf{g}\right],\label{eq:derfinicao-F2}
\end{eqnarray}
in which $\beta_{k,k+1}^{i,j}$, $\nabla_{i}W(\beta_{k,k+1}^{i,j})$,
and $\nabla_{i}W(\beta_{k-1,k}^{i,j})$ are defined by equations (\ref{eq:Definition-of-point}),
(\ref{eq:derivW00}), and (\ref{eq:Definition-of-two-points}), respectively.

To prove that $\mathbf{F}$ is contraction, we should find a constant
$c\in[0,1)$ satisfying:
\[
d\left(\mathbf{F}\left((\qb_{k+1})_{1},\qb_{k},\qb_{k-1}\right),\mathbf{F}\left((\qb_{k+1})_{2},\qb_{k},\qb_{k-1}\right)\right)\leq c\cdot d\left((\qb_{k+1})_{1},(\qb_{k+1})_{2}\right),
\]
where $(\qb_{k+1})_{1},(\qb_{k+1})_{2}\in\mathbb{R}^{3M}$ and $d:\mathbb{R}^{3M}\times\mathbb{R}^{3M}\rightarrow\mathbb{R}^{+}$
is a suitable distance function, in this case:

\[
d\left(\mathbf{q},\mathbf{r}\right)=\max\left\{ \Biggl\|\mathbf{q}^{i}-\mathbf{r}^{i}\Biggl\|_{2},\quad i=1,2,\ldots,M\right\} ,
\]
where $\|\cdot\|_{2}$ means 2-norm.

From the traditional calculus we know that if $\boldsymbol{f}:U\subset\R^{n}\to\R^{n}$
is differentiable with $\|\partial\boldsymbol{f}/\partial\boldsymbol{x}\|\leq M_{1}$
for any $\boldsymbol{x}\in U$ then \textbf{$f$} is Lipschitz; that
means, $\|\boldsymbol{f}(\boldsymbol{y})-\boldsymbol{f}(\boldsymbol{x})\|\leq M_{1}\|\boldsymbol{y}-\boldsymbol{x}\|$,
$\forall\boldsymbol{x},\boldsymbol{y}\in U$. If we show that $\mathbf{F}$
is Lipschitz then our problem turns out in finding conditions to assure
that $0\leq M_{1}<1$ in order to apply a fixed point iterative method
to approximate the solution of the equation (\ref{eq:ELdiscreto1}).
Moreover, the derivative of $\boldsymbol{F}^{i}$ respect to $\mathbf{q}_{k+1}^{s}$
is:
\begin{equation}
\dparcial{\boldsymbol{F}^{i}}{\mathbf{q}_{k+1}^{s}}=(\Delta t)^{2}(1-\alpha)\dparcial{\boldsymbol{H}_{g}^{i}}{\mathbf{q}_{k+1}^{s}}(\mathbf{q}_{k+1},\mathbf{q}_{k})\label{eq:derivative-F-respect-kplus1}
\end{equation}
where $\boldsymbol{H}_{g}^{i}\left(\qb_{k+1},\qb_{k}\right)$ is define
by equation (\ref{eq:derfinicao-F1}). To demonstrate that expression
(\ref{eq:derivative-F-respect-kplus1}) is bounded, we need to prove
that the density and pressure are bounded fields. The density $\rho_{k,k+1}^{i}=\rho((1-\alpha)\qb_{k}^{i}+\alpha\qb_{k+1}^{i})$
is computed by expression (\ref{eq:formula-gilson00}). Once the kernel
$W$ is bounded ($W(R)\leq\ell$ in expression (\ref{eq:Gaussian-Kernel})),
we can write:

\begin{equation}
\rho_{k,k+1}^{i}=\sum_{j=1}^{M}m_{j}W(\cdot)\leq\sum_{j=1}^{M}m_{j}\ell=c_{\rho},\label{eq:limit-for-density}
\end{equation}
where $c_{\rho}$ is a constant that must satisfies $c_{\rho}<\rho_{0}$
in order to get a pressure from the state equation (\ref{eq:state-equation-sph})  with physical
sense. According to equation (\ref{eq:state-equation-sph}), the pressure $p_{k,k+1}^{i}=p((1-\alpha)\qb_{k}^{i}+\alpha\qb_{k+1}^{i})$
is given by:
\begin{equation}
p_{k,k+1}^{i}=B\left[\left(\dfrac{\sum_{j=1}^{M}m_{j}W(\cdot)}{\rho_{0}}\right)^{7}-1\right]\leq B\left[\left(\dfrac{c_{\rho}}{\rho_{0}}\right)^{7}-1\right]=c_{p}\label{eq:limit-for-pressure}
\end{equation}
with $c_{p}$ being constant. Consequently:
\[
\Biggl|\dfrac{p_{k,k+1}^{i}}{(\rho_{k,k+1}^{i})^{2}}\Biggl|=\Biggl|B\left[\dfrac{(\rho_{k,k+1}^{i})^{5}}{(\rho_{0})^{7}}-\dfrac{1}{(\rho_{k,k+1}^{i})^{2}}\right]\Biggl|
\]
\begin{equation}
\leq\Biggl|\dfrac{B(\rho_{k,k+1}^{i})^{5}}{(\rho_{0})^{7}}\Biggl|+\Biggl|\dfrac{B}{(\rho_{k,k+1}^{i})^{2}}\Biggl|\leq\Biggl|\dfrac{B(c_{\rho})^{5}}{(\rho_{0})^{7}}\Biggl|+\Biggl|\dfrac{B}{(m_{i}\ell)^{2}}\Biggl|=k_{i}.\label{eq:limit-for-combination}
\end{equation}

Therefore, we can now seek for a bound for expression:

\[
\dparcial{\boldsymbol{H}_{g}^{i}}{\mathbf{q}_{k+1}^{s}}(\mathbf{q}_{k+1},\mathbf{q}_{k})=-\sum_{j=1}^{M}m_{j}\frac{\partial}{\partial\mathbf{q}_{k+1}^{s}}\left(\dfrac{p_{k,k+1}^{i}}{\left(\rho_{k,k+1}^{i}\right)^{2}}+\dfrac{p_{k,k+1}^{j}}{\left(\rho_{k,k+1}^{j}\right)^{2}}\right)\nabla_{i}W(\beta_{k,k+1}^{i,j})
\]
\begin{equation}
-\sum_{j=1}^{M}m_{j}\left(\dfrac{p_{k,k+1}^{i}}{\left(\rho_{k,k+1}^{i}\right)^{2}}+\dfrac{p_{k,k+1}^{j}}{\left(\rho_{k,k+1}^{j}\right)^{2}}\right)\frac{\partial}{\partial\mathbf{q}_{k+1}^{s}}\nabla_{i}W(\beta_{k,k+1}^{i,j}),\label{eq:derivative-Hg-resp-kplus1}
\end{equation}
where, according to equation (\ref{eq:derivW00}):
\begin{equation}
\nabla_{i}W\left(\beta_{k,k+1}^{i,j}\right)=-2\ell e^{-R^{2}}\left[(1-\alpha)\mathbf{q}_{k}^{i}+\alpha\mathbf{q}_{k+1}^{i}-\left((1-\alpha)\mathbf{q}_{k}^{j}+\alpha\mathbf{q}_{k+1}^{j}\right)\right],\label{eq:gradiwbeta}
\end{equation}
which is bounded in the considered domain.

The analysis of the first term in equation (\ref{eq:derivative-Hg-resp-kplus1})
can be made by considering the general term:
\[
\frac{\partial}{\partial\mathbf{q}_{k+1}^{s}}\left(\dfrac{p_{i}^{k,k+1}}{\left(\rho_{i}^{k,k+1}\right)^{2}}\right)=\frac{\partial}{\partial\mathbf{q}_{k+1}^{s}}\left(B\left[\frac{1}{\left(\rho_{0}\right)^{2}}\left(\frac{\rho_{k,k+1}^{i}}{\rho_{0}}\right)^{5}-\frac{1}{\left(\rho_{i}^{k,k+1}\right)^{2}}\right]\right)
\]
\[
=\frac{5B}{\left(\rho_{0}\right)^{2}}\left(\frac{\sum_{j=1}^{M}m_{j}W\left(\beta_{k,k+1}^{i,j}\right)}{\rho_{0}}\right)^{4}\times\frac{\left(\sum_{j=1}^{M}m_{j}\frac{\partial}{\partial\mathbf{q}_{k+1}^{s}}W\left(\beta_{k,k+1}^{i,j}\right)\right)}{\rho_{0}}
\]
\begin{equation}
+2B\frac{\sum_{j=1}^{M}m_{j}\frac{\partial}{\partial\mathbf{q}_{k+1}^{s}}W\left(\beta_{k,k+1}^{i,j}\right)}{\left(\rho_{i}^{k,k+1}\right)^{3}}.\label{eq:definicao-funcao-F4}
\end{equation}

Expression (\ref{eq:definicao-funcao-F4}) depends on basic operations
involving the density $\rho$, which is bounded by $c_{\rho}$, the
Gaussian kernel $W$ (expression (\ref{eq:Gaussian-Kernel})), and
its first order derivatives which are also bounded. The second term
of equation (\ref{eq:derivative-Hg-resp-kplus1}) includes derivatives
of second order of the Gaussian kernel $W$:
\[
\frac{\partial}{\partial\mathbf{q}_{k+1}^{s}}\nabla_{i}W(\beta_{k,k+1}^{i,j})=\frac{\partial}{\partial\mathbf{q}_{k+1}^{s}}\left(-2\ell e^{-R^{2}}\left[(1-\alpha)\mathbf{q}_{k}^{i}+\alpha\mathbf{q}_{k+1}^{i}-\left((1-\alpha)\mathbf{q}_{k}^{j}+\alpha\mathbf{q}_{k+1}^{j}\right)\right]\right)
\]

\[
=-2\frac{\partial W\left(R\right)}{\partial\mathbf{q}_{k+1}^{s}}\left(\left[(1-\alpha)\mathbf{q}_{k}^{i}+\alpha\mathbf{q}_{k+1}^{i}-\left((1-\alpha)\mathbf{q}_{k}^{j}+\alpha\mathbf{q}_{k+1}^{j}\right)\right]\right)
\]

\begin{equation}
-2\ell e^{-R^{2}}\frac{\partial}{\partial\mathbf{q}_{k+1}^{s}}\left(\left[(1-\alpha)\mathbf{q}_{k}^{i}+\alpha\mathbf{q}_{k+1}^{i}-\left((1-\alpha)\mathbf{q}_{k}^{j}+\alpha\mathbf{q}_{k+1}^{j}\right)\right]\right)\label{eq:derivative-bound00}
\end{equation}
that is also bounded. This fact together with expression (\ref{eq:limit-for-combination})
demonstrate that the second term in equation (\ref{eq:derivative-Hg-resp-kplus1})
is also bounded. Therefore, considering these results we claim that
there is a constant $M_{2}$ such that:

\begin{equation}
\Biggl\|\dparcial{\mathbf{H}_{g}^{i}}{\mathbf{q}_{k+1}^{s}}\left(\qb_{k+1},\qb_{k}\right)\Biggl\|_{2}<M_{2}.\label{eq:limit-for-derivative-Hg}
\end{equation}

Consequently:

\begin{equation}
\Biggl\|\dparcial{\mathbf{F}^{i}}{\mathbf{q}_{k+1}^{s}}\Biggl\|_{2}=\Biggl\|(\Delta t)^{2}(1-\alpha)\dparcial{\mathbf{H}_{g}^{i}}{\mathbf{q}_{k+1}^{s}}(\mathbf{q}_{k+1},\mathbf{q}_{k})\Biggl\|_{2}\leq(\Delta t)^{2}(1-\alpha)M_{2}\equiv M_{1},\label{eq:inequality00}
\end{equation}
that means, $\mathbf{F}^{i}$ is Lipschitz. As a consequence of the
theorem above stated we can write:
\begin{equation}
\Biggl\|\mathbf{F}^{i}\left((\qb_{k+1})_{1},\qb_{k},\qb_{k-1}\right)-\mathbf{F}^{i}\left((\qb_{k+1})_{2},\qb_{k},\qb_{k-1}\right)\Biggl\|_{2}\leq M_{1}\|(\qb_{k+1})_{1}-(\qb_{k+1})_{2}\|_{2}.\label{eq:inequality01}
\end{equation}

To assure that the function $\mathbf{F}$ in expression $(\ref{eq:derfinicao-F0})$
is a contraction mapping we need to satisfy:

\[
d\left(\mathbf{F}\left((\qb_{k+1})_{1},\qb_{k},\qb_{k-1}\right),\mathbf{F}\left((\qb_{k+1})_{2},\qb_{k},\qb_{k-1}\right)\right)=
\]

\[
=\max_{i\in\left\{ 1,\ldots,M\right\} }\left\{ \Biggl\|\mathbf{F}^{i}\left((\qb_{k+1})_{1},\qb_{k},\qb_{k-1}\right)-\mathbf{F}^{i}\left((\qb_{k+1})_{2},\qb_{k},\qb_{k-1}\right)\Biggl\|_{2}\right\} .
\]

Hence, there exists a $m\in\left\{ 1,2,\ldots,M\right\} $, such that:

\[
d\left(\mathbf{F}\left((\qb_{k+1})_{1},\qb_{k},\qb_{k-1}\right),\mathbf{F}\left((\qb_{k+1})_{2},\qb_{k},\qb_{k-1}\right)\right)=\Biggl\|\mathbf{F}^{m}\left((\qb_{k+1})_{1},\qb_{k},\qb_{k-1}\right)-\mathbf{F}^{m}\left((\qb_{k+1})_{2},\qb_{k},\qb_{k-1}\right)\Biggl\|_{2},
\]

which, by using expression (\ref{eq:inequality01}), renders:

\[
d\left(\mathbf{F}\left((\qb_{k+1})_{1},\qb_{k},\qb_{k-1}\right),\mathbf{F}\left((\qb_{k+1})_{2},\qb_{k},\qb_{k-1}\right)\right)\leq M_{1}\|(\qb_{k+1})_{1}-(\qb_{k+1})_{2}\|_{2}.
\]

Therefore, we must impose that $M_{1}\leq1$, which implies:
\begin{equation}
\Delta t<\dfrac{1}{\sqrt{(1-\alpha)M_{2}}},\label{eq:final-condition-contraction}
\end{equation}
once $M_{1}$and $M_{2}$ are related by expression (\ref{eq:inequality00}).
The above equation gives the range for $\Delta t$ that allows to
apply the fixed point framework to solve the implicit integrator given
by expression (\ref{eq:ELdiscreto1}). The practical consequences
of the bound given above is application dependent and its utility
will be analysed in the experimental results.

\subsection{Implementation Details \label{subsec:Implementation}}
Before simulating SPH with expression (\ref{eq:ELdiscreto1}), we
need to add extra machinery to account for numerical/computational aspects and interaction
of particles with boundaries.
In order to add stability to the scheme defined
by expression (\ref{eq:ELdiscreto1}),
we follow \cite{Liu2003b} and include the artificial viscosity:

\begin{equation}
\Pi_{k-1,k}^{ij}=\left\{ \begin{array}{cc}
\dfrac{-2(a\varphi_{ij}c+b\varphi_{ij}^2)}{\rho^{i}_{k-1,k}+\rho^{j}_{k-1,k}}, & \mathbf{v}_{k-1,k}^{ij}\cdot\mathbf{x}_{k}^{ij}<0\\
0, & \mathbf{v}_{k-1,k}^{ij}\cdot\mathbf{x}_{k}^{ij}\geq0
\end{array}\right.\text{ and }\varphi_{ij}=\dfrac{(\mathbf{v}_{k-1,k}^{ij}\cdot\mathbf{x}_{k}^{ij})h}{\left(r_{k}^{ij}\right)^{2}+0.01h^{2}}\label{viscosidade}
\end{equation}
where:
\begin{equation}
\mathbf{v}_{k-1,k}^{ij}=\mathbf{v}_{k-1,k}^{i}-\mathbf{v}_{k-1,k}^{j}, \label{new-equation}
\end{equation}
with:
\begin{equation}
\mathbf{v}_{k-1,k}^{i}=\frac{\mathbf{q}_{k}^{i}-\mathbf{q}_{k-1}^{i}}{\Delta t}, \label{eq:discrete-velocity}
\end{equation}
and $\mathbf{x}_{k}^{ij}=\mathbf{q}_{k}^{i}-\mathbf{q}_{k}^{j}$, $r_{k}^{ij}=\|\mathbf{x}_{k}^{ij}\|$ ($\| \cdot \|$ means the Euclidean norm). The constants
$a$ and $b$ are typically set around $1$, the constants $c$ and $h$ represent the speed of sound and smoothing length, respectively,

Besides, we shall consider repulsive boundary forces to prevent interior particles to penetrate the frontiers of the domain. In
this work this is implemented using boundary particles that do not
move but interact with fluid particles \cite{Liu2003b}. Specifically, if
a boundary particle $\mathbf{q}^{g}$ is in the neighborhood of
a real particle $\mathbf{q}_{k}^{i}$ that is approaching the boundary,
then the force:
\begin{equation}
\mathbf{\Gamma}_{k}^{ig}=\left\{ \begin{array}{cc}
D\left[\left(\dfrac{r_{0}}{r_{k}^{ig}}\right)^{n_{1}}-\left(\dfrac{r_{0}}{r_{k}^{ig}}\right)^{n_{2}}\right] \dfrac{\mathbf{x}_{k}^{ig}}{r_{k}^{ig}},\quad if & \dfrac{r_{0}}{r_{k}^{ig}} \geq 1\\
\\
\mathbf{0},\quad if  & \dfrac{r_{0}}{r_{k}^{ig}} < 1
\end{array}\right.\label{bordo}
\end{equation}
is applied pairwisely along the centerline of these two particles,
where $n_{1}=12$, $n_{2}=4$, $\mathbf{x}_{k}^{ig}=\mathbf{q}_{k}^{i}-\mathbf{q}^{g}$, and $r_{k}^{ig}=\| \mathbf{x}_{k}^{ig} \|$,
and $r_{0}$ is usually selected close to the initial particles spacing.
The parameter $D$ is problem dependent and its value should be chosen
with the same order of the square of the largest velocity.

If we add expressions (\ref{viscosidade}) and (\ref{bordo}) to the right hand side of equation (\ref{verlet}) then we can define:
\begin{equation}
\mathbf{a}^{i}=-\sum_{j}m_{j}\left(\dfrac{p(\qb_{k}^{i})}{\left(\rho(\qb_{k}^{i})\right)^{2}}+\dfrac{p(\qb_{k}^{j})}{\left(\rho(\qb_{k}^{j})\right)^{2}}\right)\nabla_{i} W(\qb_{k}^{i}-\qb_{k}^{j})+\sum_j\Pi_{k-1,k}^{ij}\nabla_i W(\qb_{k}^{i}-\qb_{k}^{j})+\mathbf{\Gamma}_{k}^{ig}+\mathbf{g},\label{eq:acceleration-verlet}
\end{equation}
and compute the solution
$\qb_{k+1}^{i}$ using the iterative procedure:

\begin{eqnarray}
\vb^{i}\left(k-\frac{1}{2}\Delta t\right)&=&\dfrac{\qb_{k}^{i}-\qb_{k-1}^{i}}{\Delta t},\label{eq:Verlet-Explicit-a}\\
\vb^{i}\left(k+\frac{1}{2}\Delta t\right)&=&\vb^{i}(k-\frac{1}{2}\Delta t)+\Delta t\mathbf{a}^{i},\\ \label{eq:Verlet-Explicit-b}
\mathbf{q}_{k+1}^{i}&=&\mathbf{q}_{k}^{i}+\Delta t\vb^{i}\left(k+\frac{1}{2}\Delta t\right) \label{eq:Verlet-Explicit-c}
\end{eqnarray}
for $k=1,2,\ldots, N$. If $k=0$ in equation (\ref{eq:Verlet-Explicit-a}) then we set $\mathbf{v}^{i}\left(-1/2\right)=v_{i}\left(0\right)-(1/2)\Delta t\mathbf{g}$.
Expressions (\ref{eq:Verlet-Explicit-a})-(\ref{eq:Verlet-Explicit-c})
defines the traditional Verlet (or Leapfrog) algorithm in the SPH
literature \cite{Violeau2012Fluid,Liu2003b}.

Moreover, to include the effects of viscous, and boundary interaction
in the implicit SPH model defined by expression (\ref{eq:ELdiscreto1}), without
creating asymmetric effects, we propose in this work the following
scheme:
\begin{equation}
\dfrac{\qb_{k+1}^{i}-2\qb_{k}^{i}+\qb_{k-1}^{i}}{\Delta t}=\Delta t(1-\alpha)\mathbf{a}_{1}^{i}+\Delta t\alpha\mathbf{a}_{2}^{i}\label{eq:esquema-implicito00}
\end{equation}
where:
\[
\mathbf{a}_{1}^{i}=-\sum_{j}m_{j}\left(\dfrac{p_{k,k+1}^{i}}{\left(\rho_{k,k+1}^{i}\right)^{2}}+\dfrac{p_{k,k+1}^{j}}{\left(\rho_{k,k+1}^{j}\right)^{2}}\right)\nabla_{i}W(\mathbf{\beta}_{k,k+1}^{i,j})
\]

\begin{equation}
+\sum_{j}m_{j}\Pi_{k,k+1}^{ij}\nabla_{i}W(\mathbf{\beta}_{k,k+1}^{i,j})+\mathbf{\Gamma}_{k}^{ig}+\mathbf{g},\label{eq:formula-a1}
\end{equation}

\[
\mathbf{a}_{2}^{i}=-\sum_{j}m_{j}\left(\dfrac{p_{k-1,k}^{i}}{\left(\rho_{k-1,k}^{i}\right)^{2}}+\dfrac{p_{k-1,k}^{j}}{\left(\rho_{k-1,k}^{j}\right)^{2}}\right)\nabla_{i}W(\mathbf{\beta}_{k-1,k}^{i,j})
\]

\begin{equation}
+\sum_{j}m_{j}\Pi_{k-1,k}^{ij}\nabla_{i}W(\mathbf{\beta}_{k-1,k}^{i,j})+\mathbf{\Gamma}_{k}^{ig}+\mathbf{g}. \label{eq:formula-a2}
\end{equation}

The direct computation of the fluid density $\rho$ using equation
(\ref{eq:density-sph00}) is not recommended due to computational and
precision problems. Therefore, following \cite{Ihmsen}, we update
the density field using the continuity equation:
\begin{equation}
\dfrac{D\rho}{Dt}=-\rho\nabla\cdot\mathbf{v.}\label{eq:continuity-equation}
\end{equation}

Using equation (\ref{eq:sph-divergent}) to write the kernel version
of the right hand side of expression (\ref{eq:continuity-equation}),
and finite differences to approximate the left hand side of this expression
we get:
\[
\frac{\rho_{k,k+1}^{i}-\rho_{k-1,k}^{i}}{\triangle t}
\]
\begin{equation}
=-\sum_{j}m_{j}\mathbf{v}_{k,k+1}^{ij}\cdot\nabla_{i}W(\mathbf{\beta}_{k,k+1}^{i,j})\label{eq:update-density}
\end{equation}

\[
\frac{\rho_{k-1,k}^{i}-\rho_{k-2,k-1}^{i}}{\triangle t}
\]
\begin{equation}
=-\sum_{j}m_{j}\mathbf{v}_{k-1,k}^{ij}\cdot\nabla_{i}W(\mathbf{\beta}_{k-1,k}^{i,j})\label{eq:update-density-1}
\end{equation}
where $\mathbf{v}_{k-1,k}^{ij}$, $\mathbf{v}_{k,k+1}^{ij}$ are
given by expression (\ref{new-equation}), and $\nabla_{i}W(\mathbf{\beta}_{k,k+1}^{i,j})$, $\nabla_{i}W(\mathbf{\beta}_{k-1,k}^{i,j})$ are calculated
through equations (\ref{eq:derivW00}) and (\ref{eq:derivW003}).

Along the SPH computation, we must evaluate the kernel
$W$, or its derivatives, in the particles positions to calculate the expressions that appear. In order to
avoid unnecessary computational overload, we set a smoothing length
$h$ that prunes the support of the kernel as follows:

\begin{equation}
W\left(R\right)=\left\{ \begin{array}{cc}
\ell e^{-R^{2}},\; if\; & R\leq h\\
\\
0, & othewise.
\end{array}\right.\label{eq:gaussian-support-limited}
\end{equation}

So, given a particle $\qb^i$, we must compute the SPH expressions only inside a neighborhood $V_{i} = \{\qb^j; \; \|\qb^i - \qb^j \| \leq h \}$. In this way,
we can use a regular data structures in order to find neighbors
quickly, as usual in the SPH literature \cite{Liu2003b}. The implicit SPH procedure is summarized by the Algorithm \ref{Alg:Algorithm-SPH-IMPLI}.

\begin{algorithm}[H]
\caption{Algorithm for implicit SPH with fixed point computation.}
\label{Alg:Algorithm-SPH-IMPLI}
\begin{algorithmic}[1]
\State (a) Parameters: Particle mass $m$, number of particles $M$, number of steps $N$, number of iterations $N_{it}$, time step $\Delta t$, $\alpha$, tolerance $\varepsilon$;
\State (b) initial conditions $\mathbf{q}_{0}^{i}$, $\mathbf{q}_{1}^{i}$,
$i=1,\cdot\cdot\cdot,M$;
\While{$k\leq N$}
	\For{all particles $i$}
		\State Search for neighboring particles;
	\EndFor
	\For {$j=1,2\ldots,N_{it}$}
		\For{all particle M}
			\State Calculate pressure by equation (\ref{eq:state-equation-sph})
			\State Calculate density derivative by equation (\ref{eq:update-density})
		\EndFor
		\For{$i=1,2,\ldots,M$}
			\State Update $\rho$ with time integrator	
			\State Suppose $\mathbf{q}_{k+1,0}^{i}$; $i=1,\cdot\cdot\cdot,M$.
			\State $\qb_{k+1;j}^{i}=\mathbf{F}^{i}\left(\qb_{k+1;j-1},\qb_{k},\qb_{k-1},\alpha\right)$,
			\If{$\|\mathbf{q}_{k+1;j}^{i}-\mathbf{q}_{k+1;j-1}^{i}\|\leq \varepsilon$}
				\State $\mathbf{q}_{k+1}^{i}\leftarrow\mathbf{q}_{k+1;j}^{i}$
				\State{stop}
			\EndIf
		\EndFor
	\EndFor		
\EndWhile
\end{algorithmic}
\end{algorithm}

\begin{algorithm}
\begin{algorithmic}[1]
\label{proc:Comp-F}
\Procedure{Compute $\mathbf{F}^{i}$}{$\qb_{k+1;j-1},\qb_{k},\qb_{k-1},\alpha$}
    \State Calculate artificial viscosity $\Pi_{k,k+1}^{ij}$ and boundary forces $\mathbf{\Gamma}_{k}^{ig}$ using expressions (\ref{viscosidade}) and
    (\ref{bordo});
    \State Compute $\mathbf{a}_{1}^{i}$ and $\mathbf{a}_{2}^{i}$ through the expressions
    (\ref{eq:formula-a1})-(\ref{eq:formula-a2})
    \State Evaluate and return:
    $
    \mathbf{F}^{i}\left(\qb_{k+1;j-1},\qb_{k},\qb_{k-1}, \alpha \right)=\left(\Delta t\right)^{2}\left((1-\alpha)\mathbf{a}_{1}^{i}+\alpha\mathbf{a}_{2}^{i}\right)+2\qb_{k}^{i}-\qb_{k-1}^{i}.$
\EndProcedure
\end{algorithmic}
\end{algorithm}

In Algorithm \ref{Alg:Algorithm-SPH-IMPLI}, we follow the idea of section \ref{subsec:Fixed-Point-SPH} and compute the position of the particle $i$ at time $t=k+1$ through
an iteration scheme based on the fixed point method. Hence, we guess an initial value for $\mathbf{q}_{k+1}^{i}$, denoted by  $\mathbf{q}_{k+1,0}^{i}$ in line $14$ of
Algorithm \ref{Alg:Algorithm-SPH-IMPLI}, which is updated
in each iteration $j$ of the successive approximations, computed in line $15$ of Algorithm \ref{Alg:Algorithm-SPH-IMPLI}, until the stopping criterion
in line $16$ is achieved. The function $\mathbf{F}^{i}$ in line $15$ is implemented following the procedure  $COMPUTE\: \mathbf{F}(\qb_{k+1;j-1},\qb_{k},\qb_{k-1},\alpha)$ just bellow the Algorithm \ref{Alg:Algorithm-SPH-IMPLI}.

\section{Computational Experiments\label{sec:Computational-Experiments}}

In this section we test the time integration scheme computed by Algorithm
\ref{Alg:Algorithm-SPH-IMPLI}. In these experiments we highlight aspects
of the fixed point iteration procedure, comparison with the Verlet
(expressions (\ref{eq:Verlet-Explicit-a})-(\ref{eq:Verlet-Explicit-c})) and momentum conservation.

We use the dam breaking simulation as the numerical example to test
the evolution of the SPH system computed by the time integration scheme
in expression (\ref{eq:esquema-implicito00}) for $\alpha=0.5$. Although
idealized, the dam breaking configuration contains information that
allow engineers to know what will happen if a dam fails and how
to set up numerical models to test it. For SPH purposes the dam breaking
set up is interesting to test the numerical stability and balance
of internal forces in the fluid.

In the computational experiments performed we use $M = 1682$ SPH particles, each one with mass $0.025 kg$. The smoothing length and the parameter $\ell$ in expression (\ref{eq:gaussian-support-limited}) are set to $h=0.072$ and $\ell=9,47$. The rest density and gravitational field intensity are given by $\rho0 = 1000.0  kg/m^3$ and $g = 9.8 m/s^2$, respectively.
The values for the tolerance used in line 8 of the Algorithm \ref{Alg:Algorithm-SPH-IMPLI} is $\varepsilon = 0.001$.

\begin{figure}[H]
\center
\includegraphics[scale=0.45]{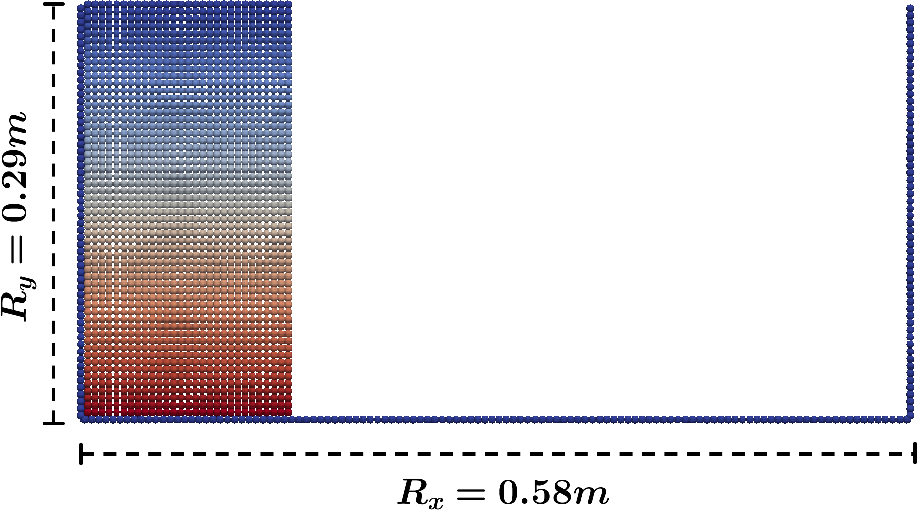}
\caption{Configuration for dam breaking simulation.}
\label{fig:dam-breaking}
\end{figure}

The computational domain, shown in Figure \ref{fig:dam-breaking},
is a rectangular region with dimensions $R_{x}=0.58 m$ and $R_{y}=0.29 m$. The initial dam, shown in Figure \ref{fig:dam-breaking}, is a fluid column with width $0.145 m$ and high $0.29 m$ filled by a regular
distribution of SPH particles with $58 \times 29 $ particles.

Firstly, we should compute the bound for the time step $\Delta t$
through equation (\ref{eq:final-condition-contraction}). Due to
expression (\ref{eq:gaussian-support-limited}), the bound $c_{\rho}$ in
(\ref{eq:limit-for-density}) depends on the estimation of the number
of particles in the neighborhood $\mathcal{B}^{i}\left(h\right)=\left\{ \mathbf{q}^{j};\|\mathbf{q}^{i}\mathbf{-q}^{j}\|\leq h\right\} $ of
a generic particle $\mathbf{q}^{i}$. Considering the dimensions of
the initial dam and the number of particles, we postulate that the cardinality
of $\mathcal{B}^{i}\left(h\right)$ has the upper bound given by $card\left(\mathcal{B}^{i}\left(h\right)\right)\approx\mathcal{O}\left(10^{1}\right)$.
Therefore $\rho_{k,k+1}^{i}\leq c_{\rho}\approx\mathcal{O}(10^{4})$,
due to expression (\ref{eq:limit-for-density}). By
substituting this result in equation (\ref{eq:limit-for-pressure})
and using the fact that $B=c^2\rho_0/7$ in this expression, we obtain
$c_{p}\approx\mathcal{O}(10^{12})$.

By substituting the bounds for $c_{\rho}$, $c_{p}$ in equations (\ref{eq:derivative-Hg-resp-kplus1})-(\ref{eq:derivative-bound00}), and by computing the bounds
for the first and second kernel derivatives, we get after some
algebra that $M_{2}\approx\mathcal{O}\left(10^{9}\right)$ and, consequently,
$\Delta t\leq10^{-4}$ is enough to apply the fixed point procedure.
Therefore, if we set $\Delta t=0.0001 s$ in initialization of the Algorithm \ref{Alg:Algorithm-SPH-IMPLI}, we satisfy the condition (\ref{eq:final-condition-contraction}).

\begin{figure}[H]
\center
\subfigure[$t=0.22s$]{\includegraphics[width=5cm]{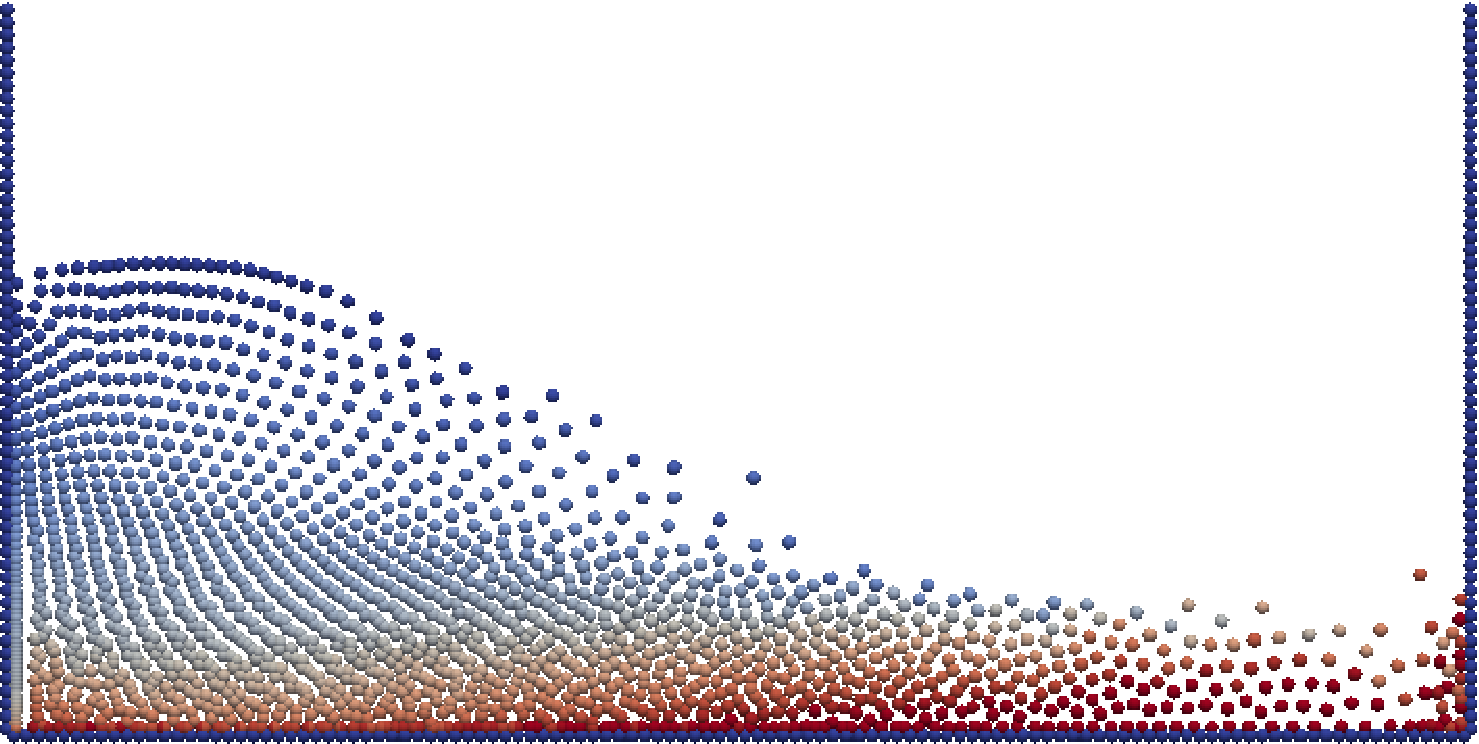}}
\quad
\subfigure[$t=0.29s$]{\includegraphics[width=5cm]{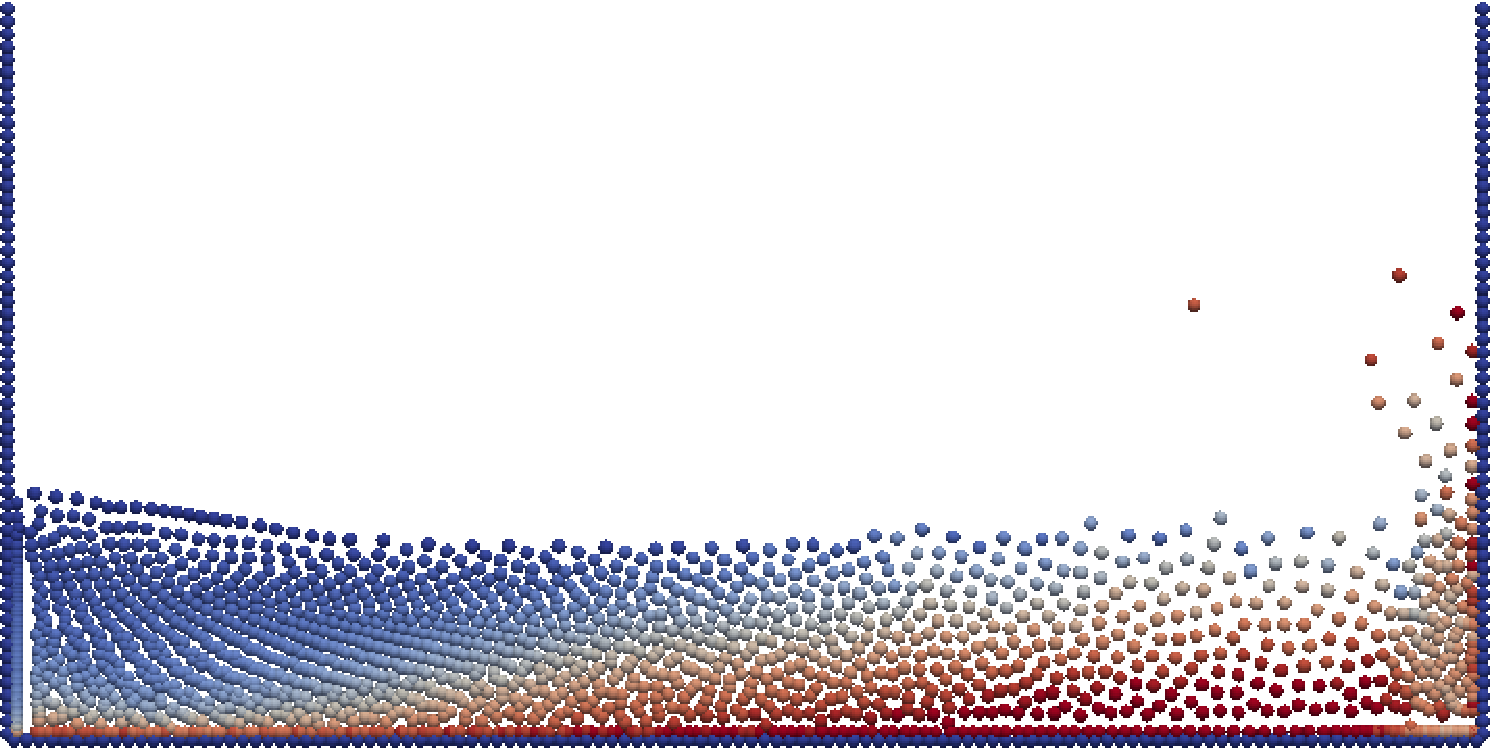}}\\
\subfigure[$t=0.35s$]{\includegraphics[width=5cm]{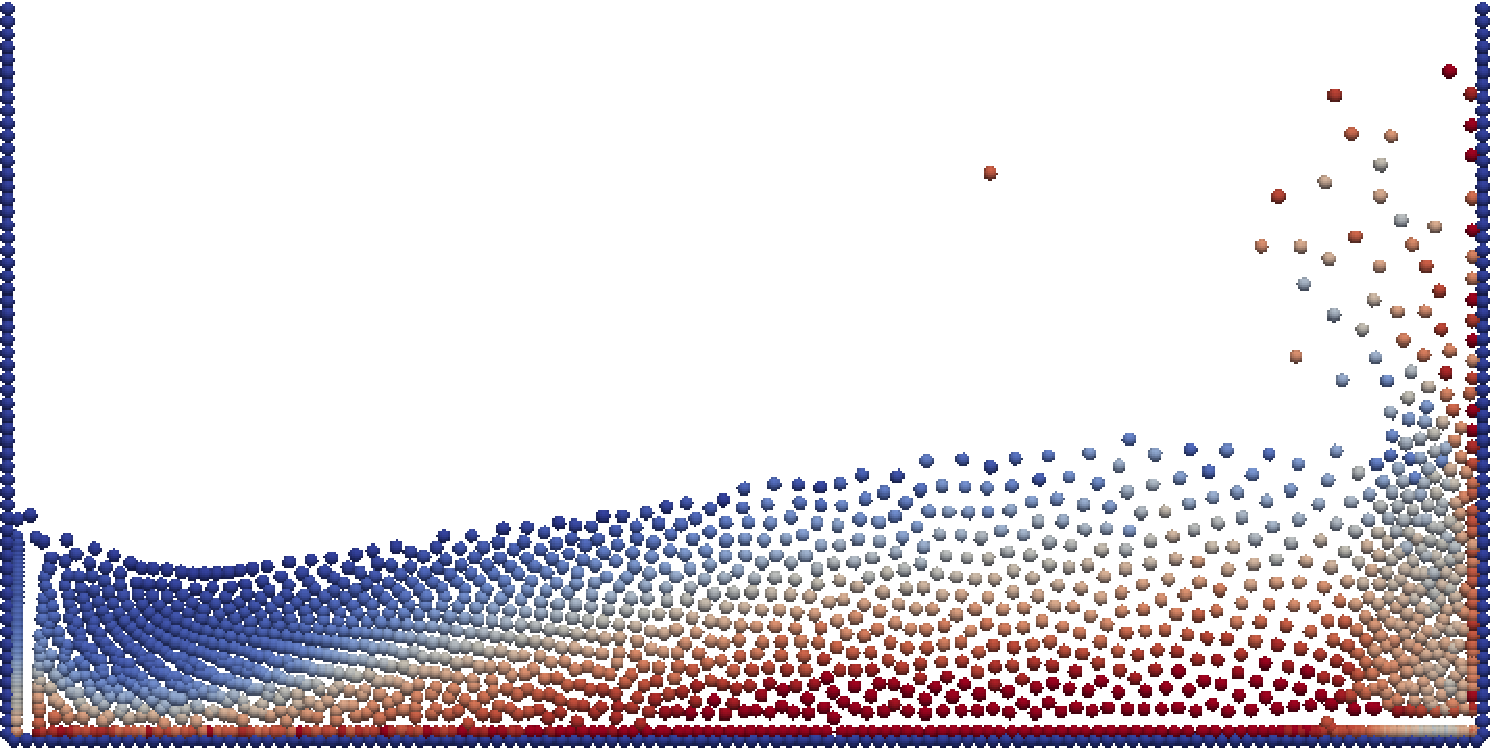}}
\quad
\subfigure[$t=0.84s$]{\includegraphics[width=5cm]{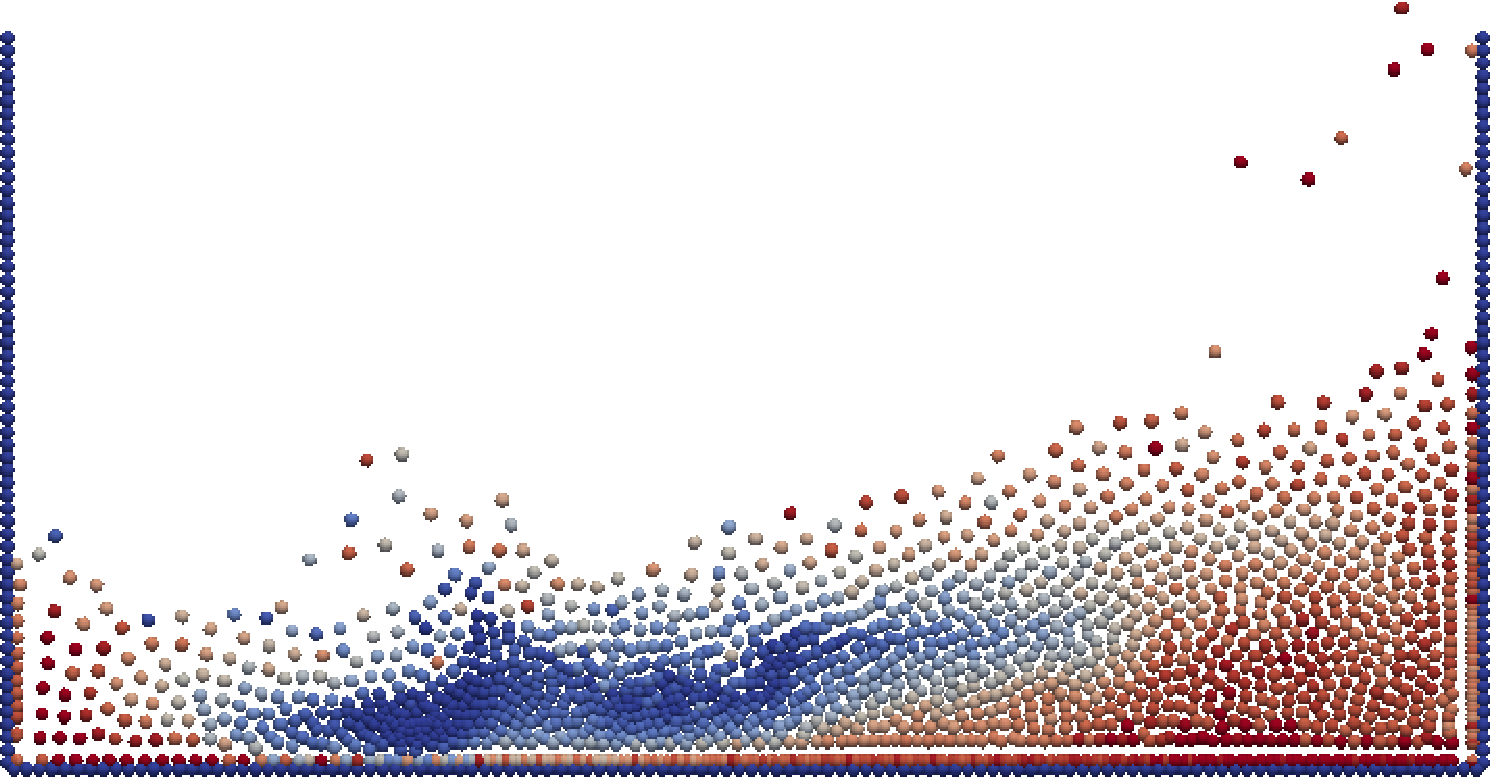}}
\caption{Dam breaking flow configuration for $\alpha=0.5$ at time steps.}
\label{fig:dam-breaking-flow-implicit}
\end{figure}

The Figures \ref{fig:dam-breaking-flow-implicit}.(a)-(d) show some snapshots of the
fluid motion with the collapse of the rectangular $2D$ dam due to
the action of the gravity field. The simulation is performed using the implicit scheme described by the Algorithm \ref{Alg:Algorithm-SPH-IMPLI}, with $\alpha=0.5$.

We also simulate the explicit scheme obtained by setting $\alpha\in\{0,1\}$
in expression (\ref{eq:esquema-implicito00}) in order
to compare a traditional SPH solution with the implicit one. With
this comparison we can visualize the differences between the implicit
and explicit simulations.

\begin{figure}[H]
\center
\subfigure[$t=0.22s$]{\includegraphics[width=5cm]{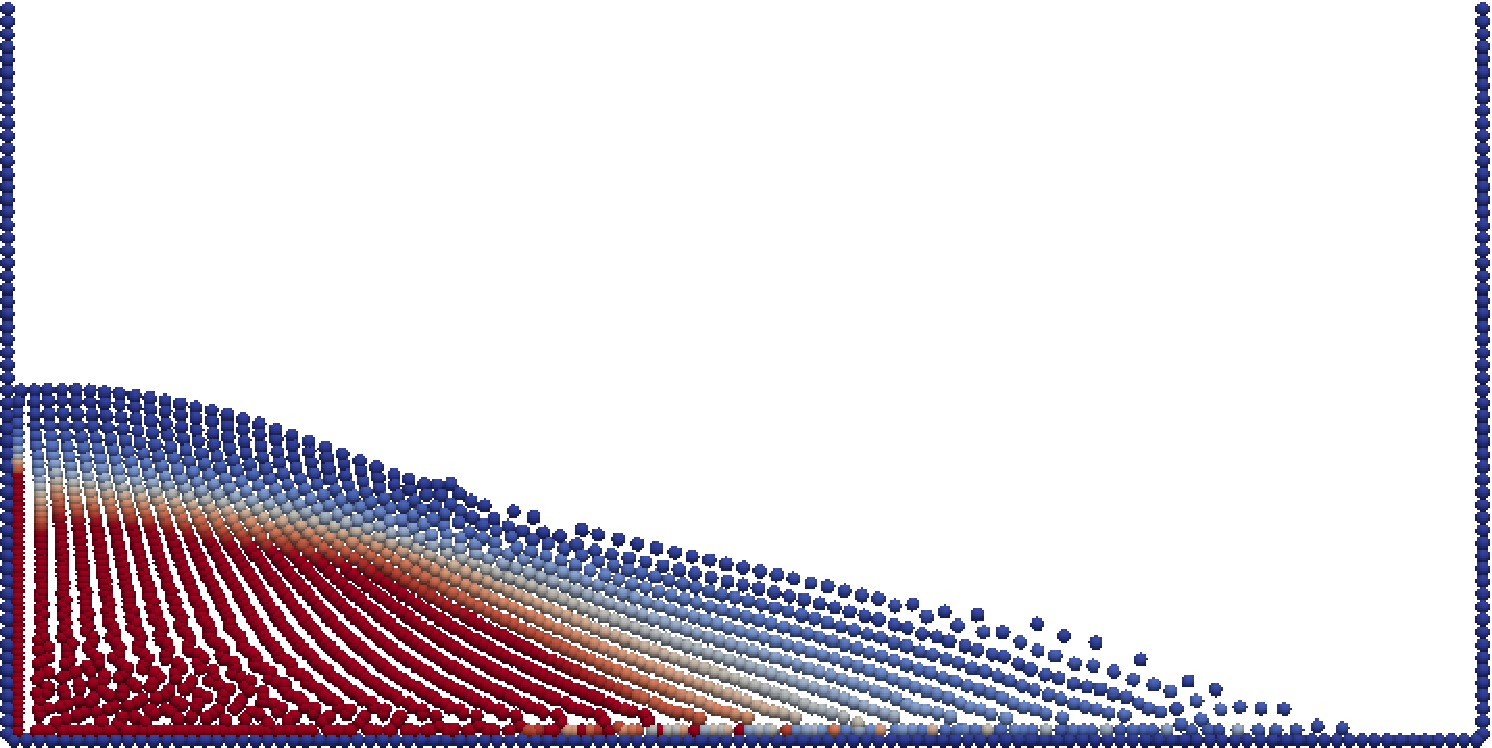}}
\quad
\subfigure[$t=0.29s$]{\includegraphics[width=5cm]{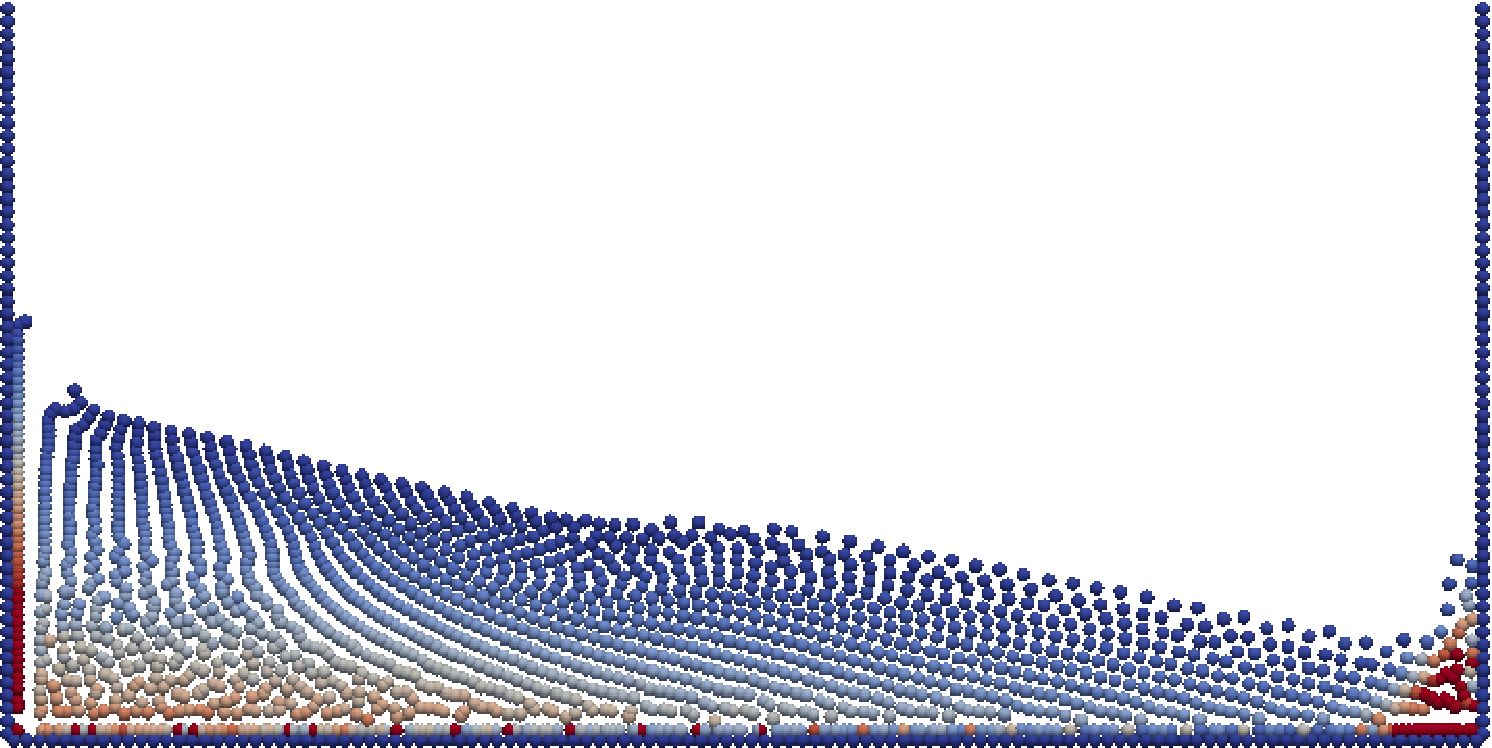}}\\
\subfigure[$t=0.35s$]{\includegraphics[width=5cm]{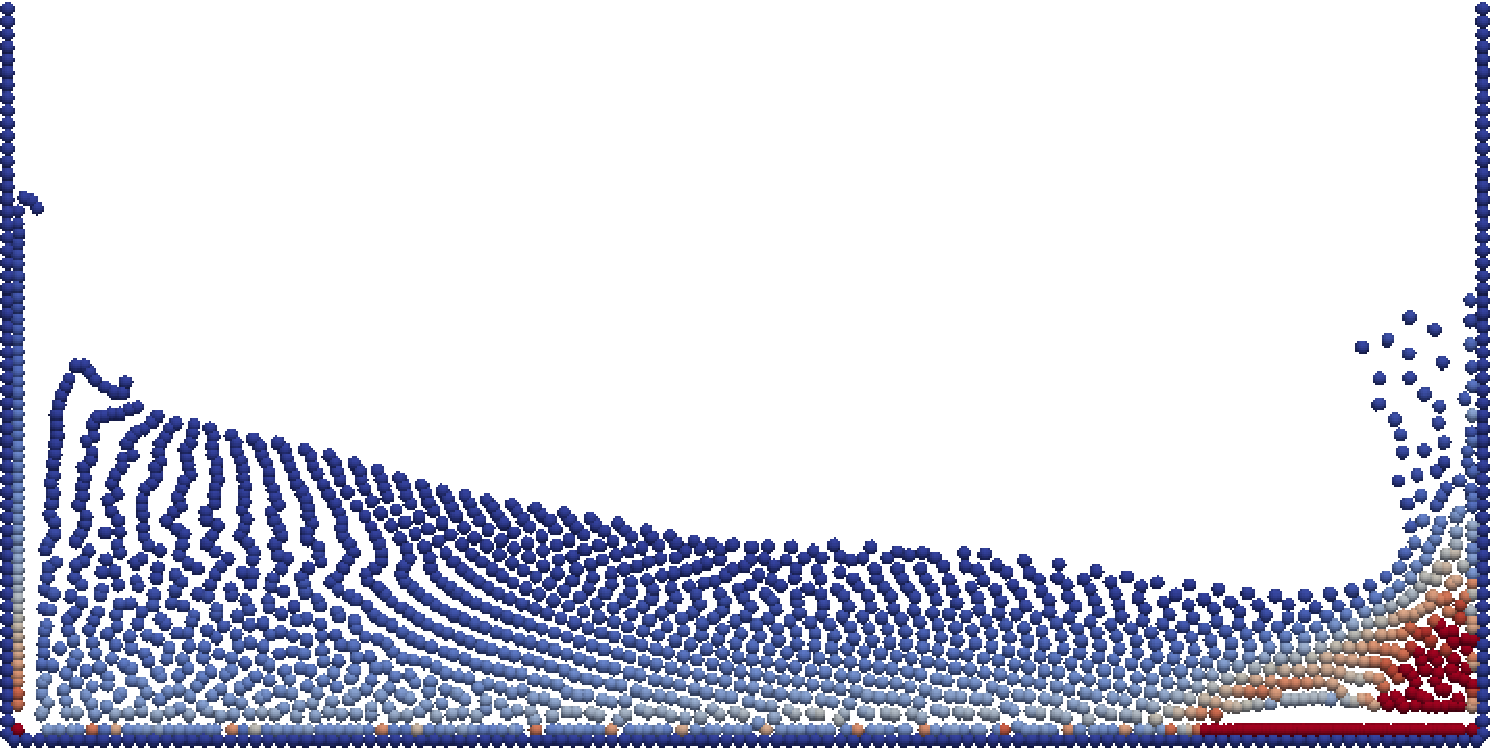}}
\quad
\subfigure[$t=0.84s$]{\includegraphics[width=5cm]{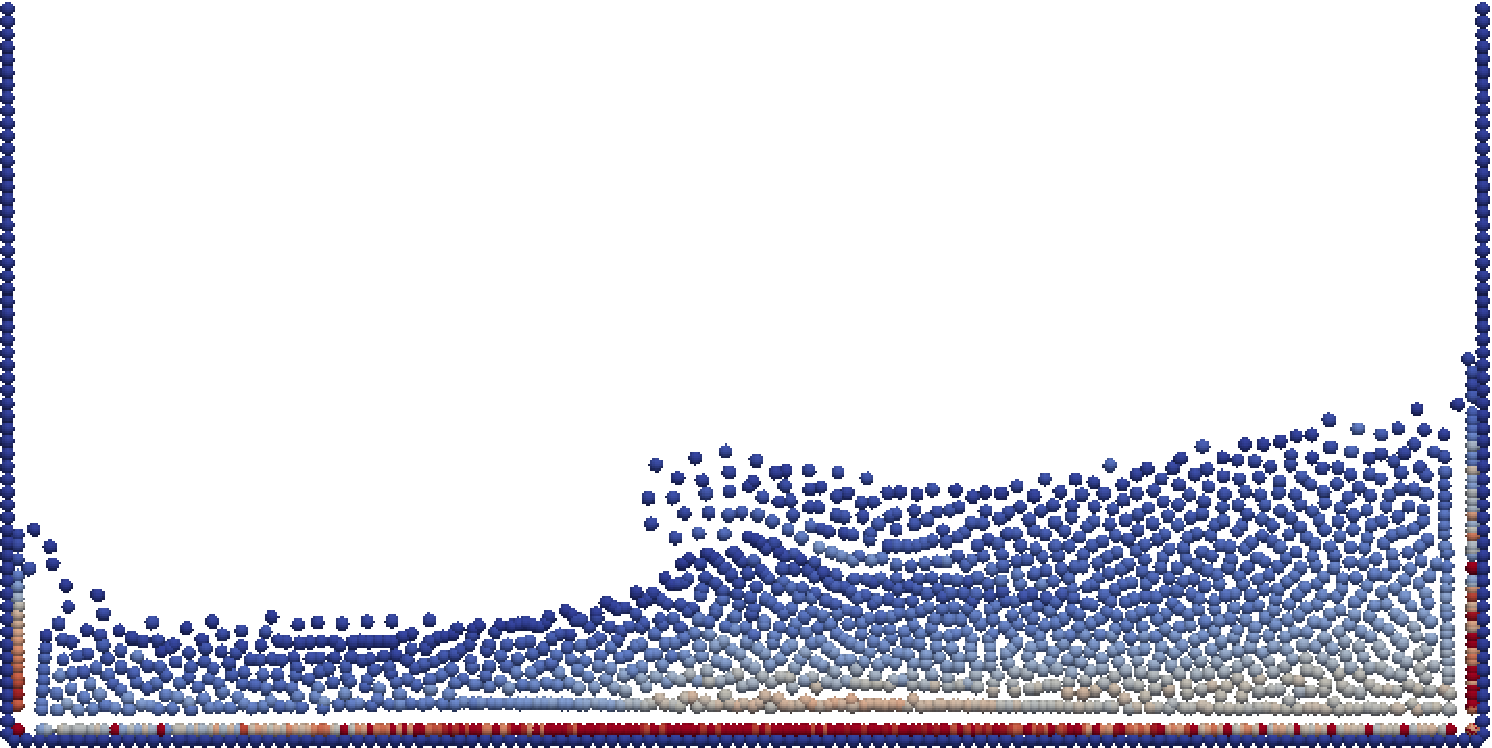}}
\caption{Dam breaking flow simulation for $\alpha\in\{0,1\}$ at time steps.}
\label{fig:dam-breaking-flow-explicit}
\end{figure}

The computation for $\alpha\in\{0,1\}$ basically follows the Algorithm \ref{Alg:Algorithm-SPH-IMPLI}
but the fixed point iterations (lines 5-13) are replaced by a direct
computation of $\mathbf{q}_{k+1}^{i}$ through equations (\ref{eq:Verlet-Explicit-a})-(\ref{eq:Verlet-Explicit-c}),
with $\mathbf{a}^{i}$ calculated by expression (\ref{eq:acceleration-verlet}).
The obtained explicit integration is computed using the same SPH and numerical parameters as before. The gravity field intensity and
the kernel are also defined like in the implicit case. The Figure \ref{fig:dam-breaking-flow-explicit}
shows four time iterations of the simulation.

When observing the results of Figures \ref{fig:dam-breaking-flow-implicit}
and \ref{fig:dam-breaking-flow-explicit} we notice that some particles go out the fluid volume, mainly in Figures \ref{fig:dam-breaking-flow-implicit}.(b)-(d).
Particles in the SPH fluid are subject to forces from neighboring
particles. Inside the fluid these inter particle forces are added and
the resultant gives the fluid motion. However, the net forces acting
on particles at the free surface may yield a resultant in the direction
of the outward surface normal, a known problem in the SPH literature
\cite{Lind-Xu-Rogers2012}, which is responsibly for the phenomena observed in Figures
\ref{fig:dam-breaking-flow-implicit}. This problem can be addressed by using an additional force field,
a surface tension, as a function of the curvature of the free surface
or even  improved versions of SPH \cite{Liu2003b,Fang-Rentschler2009}. We are not considering such approaches
in this paper.

\begin{figure}[H]
\centering
\includegraphics[width=0.35\linewidth]{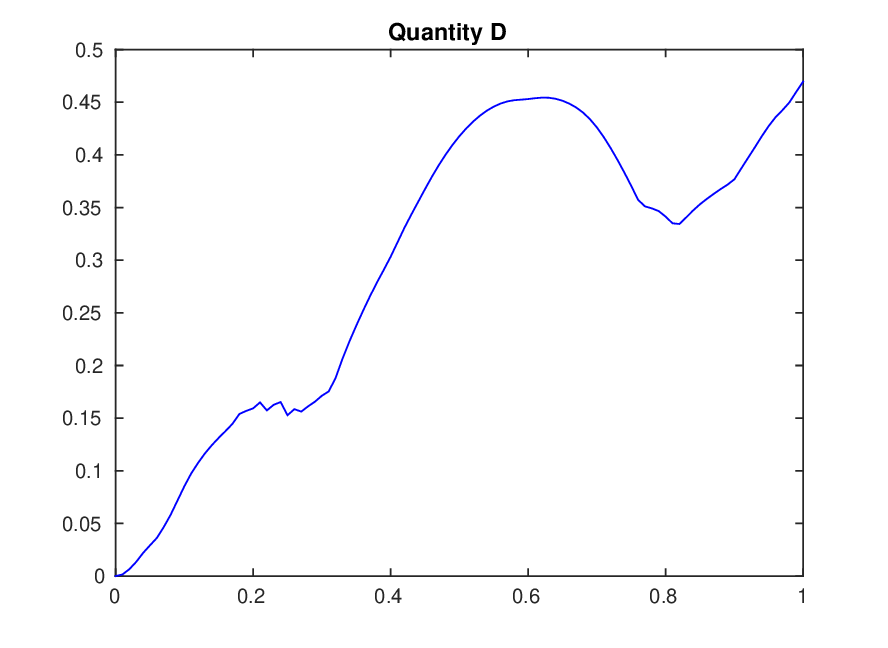}
\caption{Time evolution of expression (\ref{eq:difference-between-simulations}).}
\label{fig:quantityD}
\end{figure}

The Figure (\ref{fig:quantityD}) helps to compare the simulations for $\alpha=0.5$
and $\alpha\in\{0,1\}$. In this figure we plot the quantity $D\left(k\right)$
computed as follows:
\begin{equation}
D\left(t\right)=max_{1\leq i\leq M}\Biggl\|\mathbf{q}_{t;imp}^{i}-\mathbf{q}_{t;exp}^{i}\Biggl\|,\label{eq:difference-between-simulations}
\end{equation}
that means, given a time $t$, for each SPH particle in the implicit
simulation ($\alpha=0.5$), named $\mathbf{q}_{k;imp}^{i}$ above, we take the corresponding
SPH particles in the explicit one ($\mathbf{q}_{k;exp}^{i}$), compute
the distance between them and keep the maximum distance, which is
plotted in Figure (\ref{fig:quantityD}).

Although we can notice some oscillation of  $D\left(t\right)$ it is clear the increasing of this quantity along the simulation, which agrees with the differences observed in the snapshots of Figures \ref{fig:dam-breaking-flow-implicit} and \ref{fig:dam-breaking-flow-explicit}.

\begin{figure}[H]
\centering
\includegraphics[width=0.35\linewidth]{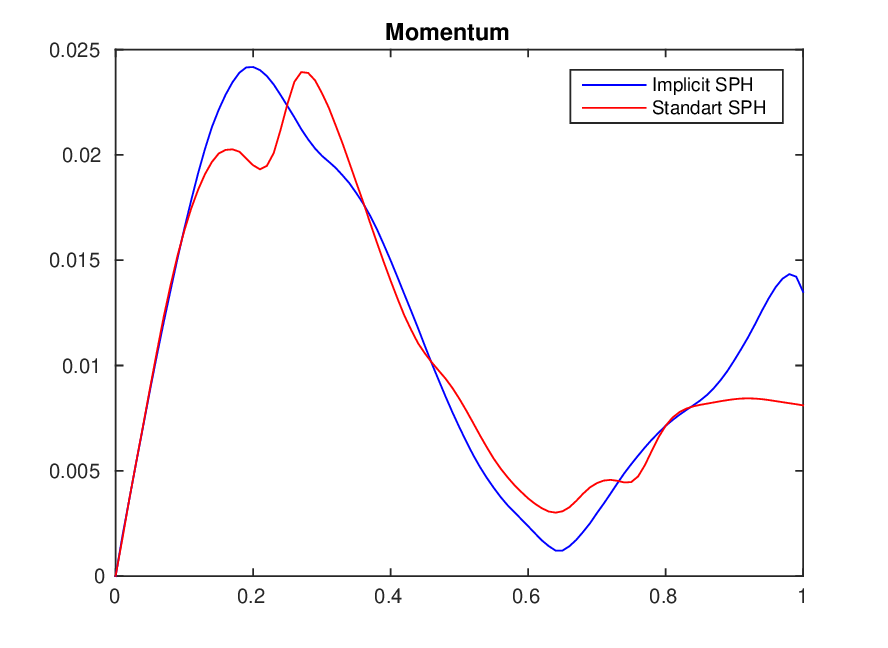}
\caption{Linear momentum for dam breaking simulation.}
\label{fig:momentum}
\end{figure}

In section \ref{subsec:Momentum-Conservation} we demonstrate that
the linear momentum of the SPH system is preserved by equation (\ref{eq:ELdiscreto1}).
However, the SPH integrator defined by expression (\ref{eq:esquema-implicito00})
includes boundary effects and the artificial viscosity. So, we shall
analyse the consequences of these extra elements for the momentum conservation.
The Figure (\ref{fig:momentum}) shows the temporal evolution of the linear momentum $L$ for
the dam breaking SPH simulation, given by:
\[
Q=\Biggl\|\frac{1}{M}\sum_{i=1}^{M}m_{i}\mathbf{v}_{k}^{i}\Biggl\|,
\]
where the velocity field is obtained by simulating the fluid using
the implicit scheme (back line) and the explicit one (red line). We
notice that linear momentum of the system oscillates and decays for both implicit and explicit schemes. It is an expected effect once the artificial viscosity dissipates the kinetic energy of the system. However, this effect is more intense in the explicit formulation as we can see in the interval $0.8s < t < 1.0s$.

\section{Conclusions and Future Works\label{sec:Conclusions-and-Future}}

The paper has presented a discrete variational formulation for fluid
simulation based on SPH. Up to the best of our knowledge, this paper
is the first one to propose such discrete setting for fluid simulation
within SPH framework. We demonstrate that the obtained variational
time integrator preserves linear momentum. Moreover, we find conditions
that support the application of fixed point theory for time integration.
Due to numerical and practical requirements, we add viscous and boundary
effects to the integration procedure. Therefore, we perform computational
experiments to analyse the consequences of these extra machinery in
the conservation property. We noticed a decreasing in the linear momentum which is less noticeable for the implicit integrator.

The midpoint numerical integration rule applied depends on a parameter
$\alpha$ which falls in the range $\left[0,1\right]$. Further works,
that analyse topological properties of the phase space when changing
the parameter $\alpha$ are currently under consideration.

\end{document}